\documentclass[trackchanges]{aastex701}
\usepackage{booktabs, mathrsfs, verbatim}

\begin{document}

\title{The First Model-Independent Upper Bound on Micro-lensing Signature of the Highest Mass Binary Black Hole Event GW231123}

\author[0009-0004-4937-4633]{Aniruddha Chakraborty}
\affiliation{Tata Institute of Fundamental Research, 
Homi Bhabha Road, Navy Nagar, Colaba, 
Mumbai 400005, India}
\email{aniruddha.chakraborty@tifr.res.in}

\author[0000-0002-3373-5236]{Suvodip Mukherjee}
\affiliation{Tata Institute of Fundamental Research, 
Homi Bhabha Road, Navy Nagar, Colaba, 
Mumbai 400005, India}
\email{suvodip@tifr.res.in}

\begin{abstract}

The recently discovered gravitational wave event, GW231123, is the most massive binary black hole merger detected to date. The inferred source masses of the event fall within the pair-instability supernova mass gap, where black holes formed directly from stellar progenitors are expected to be rare, making alternative formation scenarios for such massive black holes especially relevant. One proposed explanation is gravitational lensing, which can make the source masses to be inferred as higher than their true values.
In this work, we search for lensing signatures in GW231123, together with other O4a events, using a model-independent approach with \texttt{$\mu$-GLANCE}. The method tests residual strain for correlated features across the detector network via cross-correlation and infers lensing-induced modulations within a Bayesian framework. 
Our analysis finds no strong evidence for lensing in GW231123, but reveals a potential residual feature that could be consistent with microlensing, with a modulation amplitude of up to 0.8 at 95\% confidence. However, we find that waveform systematics for such heavy binary systems are sufficiently large to shadow the lensing signatures in short-duration signals like GW231123, preventing any definitive claim of lensing at this stage. 
We conclude that, if this event is lensed, similar lensed events will be detectable in the near future with current detector sensitivity, opening a new discovery space for lensed gravitational waves with the aid of more accurate waveform models.

\end{abstract}


\section{Introduction} 

Gravitational waves (GWs) are distortions in spacetime that propagate at the speed of light \citep{Einstein:1915ca, PhysRev.117.306, 1975ApJ...195L..51H, Cutler:1994ys}. These ripples in spacetime are originated by the time-variations in the mass-quadrupole moment of a massive system. GWs can be generated from a variety of astrophysical sources, ranging from supernovae \citep{1978Natur.274..565T, 1992AGAb....7...57M, Fryer:2001zw, Abdikamalov:2020jzn, PhysRevD.107.043008, Vartanyan:2023sxm}, compact binary mergers \citep{Chernoff:1993th, PhysRevLett.70.2984, Will:1994xx, Poisson:1995cm, Blanchet:1995fg, 1996CQGra..13..575B} to rotating neutron stars \citep{1984ApJ...278..345W, Bonazzola:1995yu, PhysRevD.55.2014}. Out of all these possibilities, the LIGO-Virgo-KAGRA (LVK) detector network \citep{KAGRA:2013rdx, LIGOScientific:2014pky, PhysRevD.111.062002, LIGO:2024kkz, PhysRevD.102.062003, PhysRevLett.123.231107, PhysRevD.93.112004, PhysRevLett.116.131103, Harry:2010zz, KAGRA:2013rdx, VIRGO:2014yos, PhysRevLett.123.231108, Virgo:2022ysc, Luck:2010rt, 2014CQGra..31v4002A, Dooley:2015fpa, KAGRA:2020tym, PhysRevD.88.043007, Somiya:2011np} has observed about 180 compact binary mergers up to GWTC-4 \citep{LIGOScientific:2018mvr, LIGOScientific:2020ibl, LIGOScientific:2021usb, KAGRA:2021vkt}, most of which are binary black hole (BBH) mergers, with a few neutron star-black hole (NSBH) and binary neutron star (BNS) systems. These sources have opened up a plethora of unexplored astrophysics: tests of fundamental physics \citep{LIGOScientific:2016lio, LIGOScientific:2018dkp, LIGOScientific:2019fpa, LIGOScientific:2020tif, LIGOScientific:2021sio}, inferences of the rates and population properties of compact binary objects \citep{LIGOScientific:2020kqk, KAGRA:2021duu}, inferences of cosmological parameters \citep{LIGOScientific:2021aug}, searches for gravitationally lensed GWs \citep{Kim:2022lex, LIGOScientific:2021izm, LIGOScientific:2023bwz} - to mention a few. 

When gravitational waves propagate near massive objects, their trajectories deviate from the straight path due to the curvature of spacetime created by the massive object \citep{1936Sci....84..506E}. This is known as gravitational lensing, which produces different effects depending on the ratio of the GW wavelength scale and the characteristic length of the lensing object. In the geometric optics regime, gravitational lensing can amplify the GW amplitude and introduce constant phase-shifts to the image GWs in a frequency-independent way \citep{1992grle.book.....S}. In the wave optics regime, gravitational lensing produces single-image, but the amplitude enhancements and phase distortions occur in a frequency-dependent way. The precise lensing effects depend on the mass distribution of the lens \citep{Takahashi:2003ix}. 

In this work, we have performed residual tests on the event GW231123\_135430 \citep{LIGOScientific:2025rsn} (which we will refer to as GW231123 from now onward) to look for any evidence of wave-optics microlensing features in the data. The most massive BBH merger so far, GW231123, has been detected on November 23rd, 2023. The primary mass lies within or just above [60 $M_{\odot}$, 130 $M_{\odot}$], and the secondary mass lies within the range. This observation challenges our key understanding of the formation of stellar black holes in this theoretical mass gap, where no stellar-origin black holes are formed. This is because, due to the pair-instability supernova \citep{2003ApJ...591..288H, Woosley:2007qp, Belczynski:2016jno, 2021ApJ...912L..31W} (PISN), the core of the massive metal-poor star explodes, leaving no black hole remnant. The observations from LVK have provided a hint towards the discovery of a PISN mass gap around 45 M$_\odot$\citep{Farmer:2019jed, Woosley:2021xba}. The spin magnitudes of the component black holes are very close to unity, making both of them very extreme cases in the Kerr (spinning) black hole scenario. Although detected with a high signal-to-noise ratio of 22.7 across the detector network, the inference of its source is heavily impacted by waveform systematics, reflecting the limitations of the current GW waveforms in the high-mass regime. The LVK collaboration, in their recent work, \citep{lvk_gwtc4_lensing} reports no statistically significant event in O4a that can be classified as gravitationally lensed candidates in the wave-optics domain. The event GW231123 although shows some support for single-image distortion lensing, the high degree of waveform uncertainty associated with GW231123 makes its interpretation as a lensed GW very challenging.

This work has been organized into the sections as follows. In section \ref{sec2}, we discuss the inferences of source properties of GW231123, and how that is impacted by waveform modeling. We also discuss the motivation behind its lensing aspects. Any astrophysical detection is confident if it relies less on the specific modeling of astronomical objects \footnote{See for example, the detection of GW150914 using "Generic Transient Search" subsection here \citep{LIGOScientific:2016aoc}. The method did not assume any specific model of the GW radiation.}. To identify wave-optics lensing without relying on the model of the mass-distribution profile of the lens, we apply cross-correlation technique between the residuals GW231123 from different detectors in section \ref{sec3}. Furthermore, to verify and characterize the lensing signature, we have applied a Bayesian technique to estimate the strength at which lensing is present in section \ref{sec4}. The other dependence on the model comes from using the best-fit GW strains of different waveforms. Thus we check the systematic error associated with waveform modeling that can arise false lensing signatures in section \ref{sec5}. In section \ref{sec6}, we show how residuals that show up in cross-correlation, can arise for a high-mass unlensed event similar to GW231123. To demonstrate this, we recover the source properties from simulated high-mass events with the help of different waveforms. This shows that with current waveform templates, short signals $\simeq 0.2s$ cannot be well-characterized, let alone tested for wave-optics lensing. We also show the result from the micro-lensing search on all O4a events in section \ref{gwtc4all}. We predict detectable lensed events with future LVK observations and how that would confirm the lensing status of GW231123 in section \ref{sec7}. 

\section{Details of the event GW231123 and its lensing status}\label{sec2}

\subsection{Brief description of GW231123}

\begin{figure}
    \centering
    \includegraphics[width=0.9\linewidth]{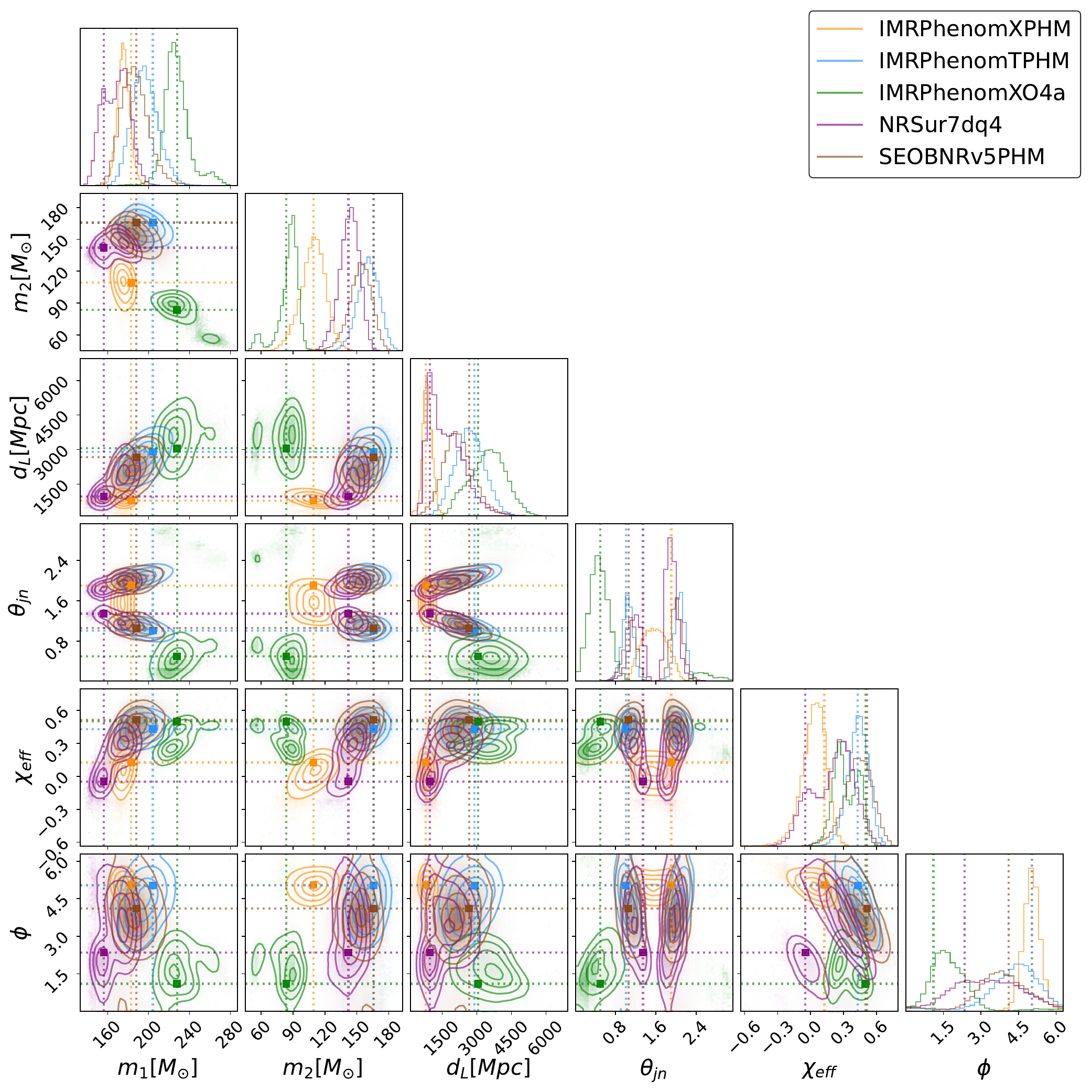}
    \caption{In this figure, we show the inference of GW source properties: detected BBH masses ($m_1$ and $m_2$), luminosity distance ($d_{\rm L}$), inclination angle ($\theta_{\rm jn}$), effective-spin ($\chi_{\rm eff}$) and coalescence angle ($\phi$). The results for five different waveform models are shown in different colors. The disagreement between the parameters recovered shows the high systematic uncertainties associated with the waveform modeling for high-mass BBH mergers. Vertical dotted lines show the maximum-likelihood values of these parameters.}
    \label{fig:source_prop}
\end{figure}

GW231123 was detected with Hanford (H1) and Livingston (L1) observatories, with a network matched-filter SNR of $\approx$ 22.7 in the detection pipelines. The signal lasted for $\approx 0.2s$ in the observable frequency band $20 - 256$Hz. A spectrogram analysis revealed a glitch in frequencies $15 - 30$Hz in the H1 detector 1s prior to the merger, but this did not have any overlap with the signal \citep{LIGOScientific:2025rsn}. 

In order to understand the GW waveform-modeling uncertainties for the high-mass black hole mergers, the inference of the GW source properties with 5 different waveform models is shown in figure \ref{fig:source_prop}. The estimated posterior distributions of the component masses, luminosity distance, inclination angle, effective spin, and coalescence phase are shown in the figure. Primary mass estimates vary in the range from 150$M_{\odot}$ to 250$M_{\odot}$ with different waveform models. Secondary mass distributions are also not well-constrained: the distributions range from 75$M_{\odot}$ to 175$M_{\odot}$. The luminosity distance posteriors also varies substantially, ranging from 1000Mpc to 4500Mpc. The angle of inclination being degenerate with the distance, also occupies different regions as the waveforms vary. Despite their differences, the effective spin parameter estimates for all these models have a significant support towards non-zero spin, with magnitudes ranging up to 0.7. By contrast, the coalescence phase estimates show very little consistency across models, spanning almost the entire range from 0 -- $2\pi$.

Given these rare properties of  this event, a few interesting studies have been performed on GW231123. The primordial origin of the source BBH has been tested \citep{DeLuca:2025fln, Yuan:2025avq}. It has also been hypothesized that the event has formed from hierarchical mergers of multiple black holes \citep{Li:2025fnf}. Population-III stars, being the progenitors of the BBH, have been hypothesized \citep{Tanikawa:2025fxw} altogether with stellar astrophysics alone being sufficient to describe the properties of the merging BBH \citep{Croon:2025gol}. The GW originating from the vibrations of cosmic strings has also been speculated \citep{Cuceu:2025fzi}. 

\subsection{Why is GW231123 important for lensing studies?}

The component masses of the BBH generating GW231123 can lie within the theoretical PISN mass gap, where no black hole is expected to form from a progenitor star. Massive metal-poor stars can form electron-positron pairs from a photon, in a process called pair production. This reduces the radiation pressure at the core, triggering a core-collapse. This enhances nuclear fusion, and the core explodes without leaving a core behind. This is pair instability supernovae (PISN). It is hypothesized that there exists a gap in black hole mass due to PISN if the black hole is formed directly from a progenitor star. The PISN mass gap spans the mass range: 60-130 $M_{\odot}$ \citep{Woosley:2021xba}. However, formation channels that do not directly rely on progenitor stars for the formation of black holes can produce black holes in the PISN mass gap. Examples of such channels include black holes of primordial origin, or black holes produced to successive mergers of less massive black holes. 

Gravitational lensing can provide an alternative explanation for the inferred source masses of GW231123, lying within the PISN gap. The strength of gravitational waves can be calculated using fundamental physics theory (namely the general theory of relativity), which means that we can infer the distance to a GW source from its strain. Given a cosmological model of the expanding universe, we can relate this distance to a redshift \footnote{The cosmic distance inferred from the luminosity of a source is known as the luminosity distance; it is given by, $d_L(z) = (1+z)c \int _0 ^z \frac{dz'}{H(z')}$, where c is the speed of light in vacuum and $H(z')$ is the Hubble parameter. In the small redshift regime, the luminosity distance is a linear function of redshift, given by, $d_L(z) = \frac{cz}{H_0}$.}. Again, general relativity theory helps us to infer the masses of the GW source in the detector frame. From inference of the masses of the source in the detector frame $m_{1d}$ and $m_{2d}$ respectively, and the redshift of the source $z$, we obtain the masses in the source frame using the relation, $m_{1s} = m_{1d} / (1+z)$ and $m_{2s} = m_{2d} / (1+z)$. When gravitational lensing magnifies the signal and makes it appear to come from a closer distance ($d_L$), it shifts the inference of the redshift ($z$) to a lower value. This makes the source masses in the source frame appear heavier. So, using the lensing hypothesis, the BBH masses can be pushed to a lower mass regime, potentially avoiding their overlap with the PISN mass gap. A strong lensing multiple image search by LVK collaboration found no strong candidate GW signal with a match to GW231123 \citep{LIGOScientific:2025rsn}.




\section{Micro-lensing search of GW231123 using Model-Independent $\mu$-GLANCE}\label{sec3}

$\mu$-\texttt{GLANCE} is a model-independent way to test the presence of wave-optics lensing effects in a GW signal. It uses the best-fit GW waveform model to calculate residuals in different detectors, which then are correlated to find any unmodelled yet common structures. The mathematical structure of this technique is described in the appendix \ref{app:1}.

Below, we mention the salient aspects of $\mu-$\texttt{GLANCE} and its model-independent nature:

\begin{enumerate}
    \item \textbf{Lensing model independence:} To claim a wave-optics feature detection, in a way irrespective of the astrophysics of the lens. It does not assume any mass-distribution model of the lensing object to detect wave-optics lensing; it searches for common structures present in the residuals on the GW detectors.
    
    \item \textbf{Waveform model independence:} By using multiple waveform models of the GW signal and observing the persistence nature of the residual features across these models, we distinguish any waveform model-dependent artifacts from truly physical signatures.
    
    \item \textbf{Confirmation of a lensing origin:} To confirm the features in the residuals are indeed from lensing, we perform a Bayesian inference in the lensing parameter space with an amplification template. The template, valid in the wave-optics regime, is based on the fact that the wave-optics lensing brings in frequency-dependent oscillatory features in the lensing amplification, and not tuned to any specific lens model.
\end{enumerate}
  
Here in this work, to find the lensing signatures from GW231123 and subsequently other GWTC-4 events, we follow the same analysis technique performed in the lensing search using GWTC-3 data in our previous works \citep{Chakraborty:2025maj}.

\begin{figure}
    \centering
    \includegraphics[width=0.9\linewidth]{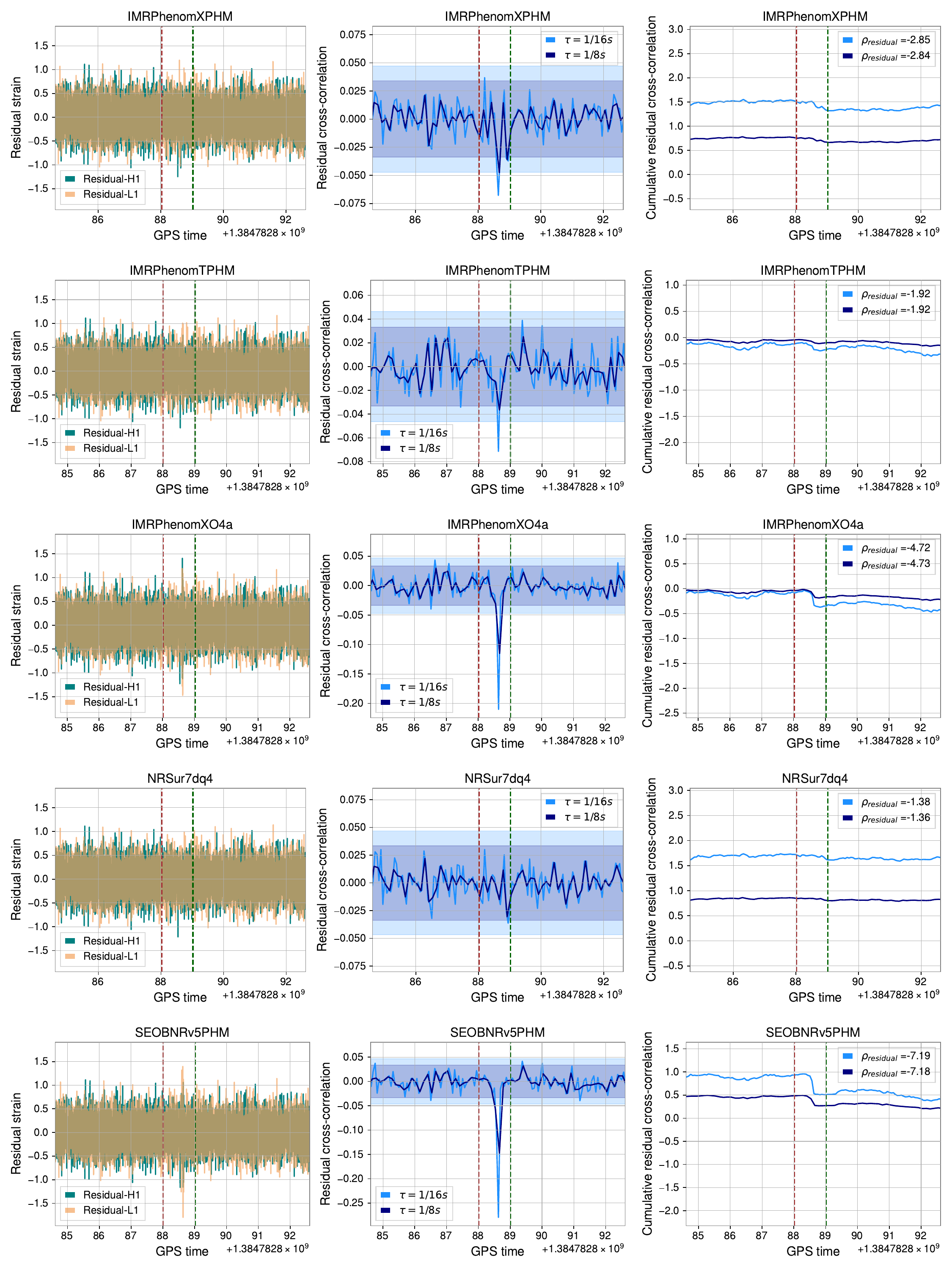}
    \caption{In this figure, along the first column, we show the residuals in detector H1 and L1. In the second column, we show residual cross-correlation for two different timescales $1/8s$ and $1/16s$, to show the variation of the residual cross-correlation with the associated timescale. In the third column we show cumulative residual cross-correlation. Each row is associated with the results given a waveform model, the model name is mentioned in the title of the figure(s). The vertical dashed green and red lines denote the start time and the end time over which the residual cross-correlation SNR is calculated. It is chosen ($1.0s$) over a sufficiently longer timescale than the in-band ($0.2s$) duration, to completely encapsulate the signal. The values of the residual cross-correlation SNR are mentioned in the third panel of each row.}
    \label{fig:res_cc}
\end{figure}

To check the presence of any correlated characteristic in the residual, we apply \texttt{$\mu$-GLANCE} on the GW231123 residuals from detectors H1 and L1. We calculate the best-fit waveform and subsequently calculate the residuals and their cross-correlation for five waveform models \texttt{IMRPhenomXPHM-SpinTaylor} \citep{Pratten:2020ceb}, \texttt{IMRPhenomTPHM} \citep{Estelles:2021gvs}, \texttt{NRSur7dq4} \citep{Varma:2019csw}, \texttt{IMRPhenomXO4a} \citep{Thompson:2023ase} and \texttt{SEOBNRv5PHM} \citep{Ramos-Buades:2023ehm} \footnote{The maximum-likelihood values of the fifteen GW source parameters are used to calculate the GW best-fit waveform. In the following step, the waveform is projected onto the detectors accounting the response functions of the detectors.}. The band-passed and whitened residuals for detectors H1 and L1 for these waveform models are shown along the first column of figure \ref{fig:res_cc}. We choose the cross-correlation timescale ($\tau_{\rm cc}$) of 0.125s and 0.0625s, shorter than (and comparable to) the in-band duration of 0.2s of the signal. The cross-correlation between the H1 and L1 residuals is shown in the second column. Due to the specific orientation of the H1 and L1 detectors on earth, any common feature in the data shows up in the negative direction \footnote{Please refer to the article \citep{lhantenna, det_antenna} for more details on the detector antenna patterns.}. The horizontal band shows the $3\sigma$ of the distribution of noise cross-correlation $N_{xx'}$ with its color representing the cross-correlation timescale as mentioned in the labels. In the third column, we show the residual SNR for a timescale ($\tau_{\rm snr}$) of 1s. The dashed vertical lines (brown and green lines) show the time-interval over which the residual cross-correlation is summed to understand how microlensing features, if present, accumulate in time.

From the third column of the figure \ref{fig:res_cc}, we show that the residual cross-correlation SNR ($\rho_{\rm residual}$) for the waveform models \texttt{IMRPhenomXPHM}, \texttt{IMRPhenomTPHM}, \texttt{IMRPhenomXO4a}, \texttt{NRSur7dq4}, and \texttt{SEOBNRv5PHM} are respectively 2.85, 1.92, 4.72, 1.38 and 7.19 respectively for cross-correlation timescale of $0.0625s$ and SNR timescale of $1s$. We show that a residual feature $\geq 3\sigma$ is supported only by \texttt{IMRPhenomXO4a} and \texttt{SEOBNRv5PHM}. However, residual cross-correlation does not inform us about the origin of the common signatures in the residuals. Therefore, to check whether the features appear due to lensing, we perform a lensing parameter estimation on the signal, using an amplification template. This is shown in the next section \ref{sec4}.

\section{Characterizing the wave-optics lensing features in a Bayesian approach}\label{sec4}

\begin{figure}
    \centering
    \includegraphics[width=0.75\linewidth]{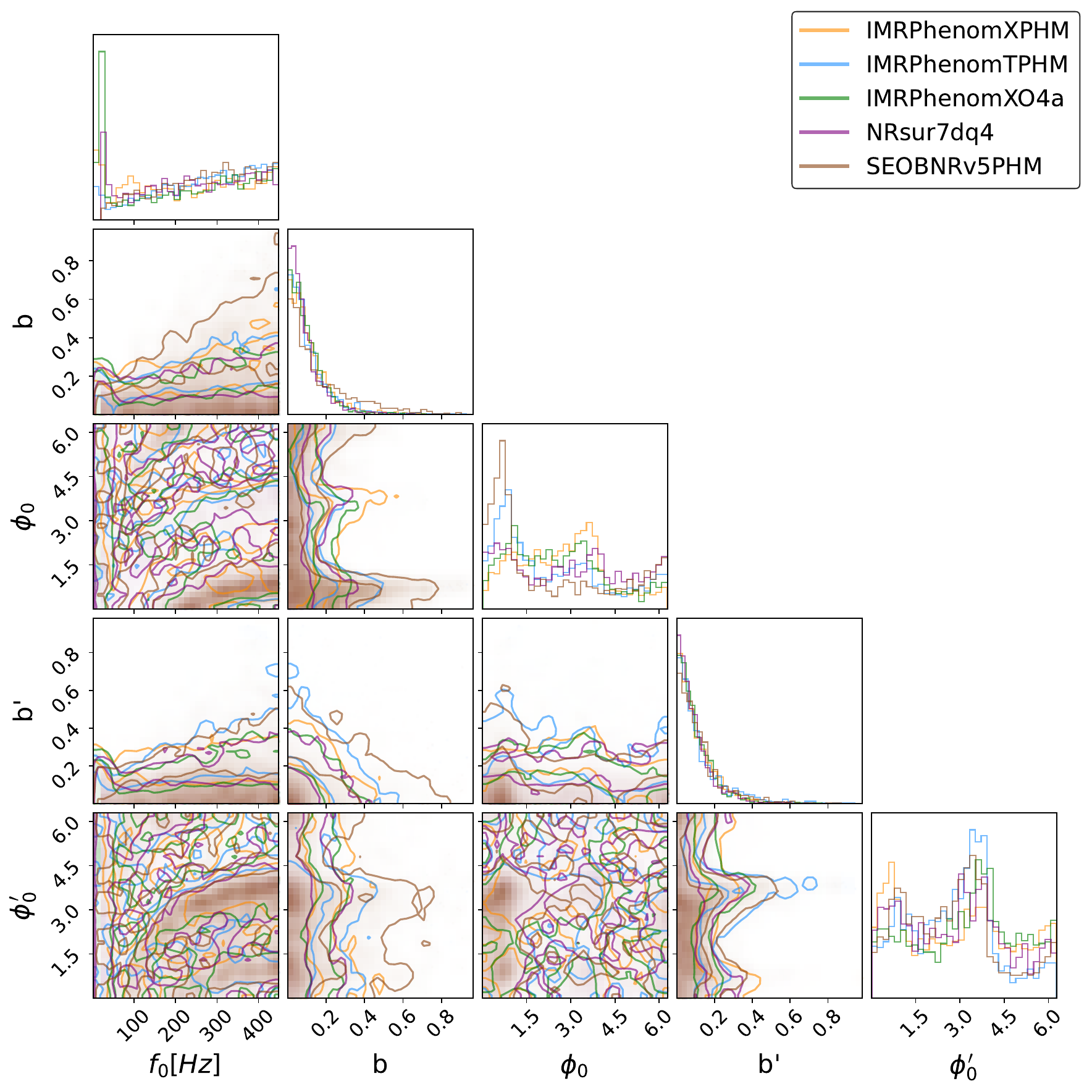}
    \caption{In this figure, we show the estimation of the lensing parameters for different waveform models. The posteriors obtained for different waveform models agree on the values of the estimated lensing parameters. The estimations of the lensing parameters show consistency with the residual cross-correlation results: higher the residual cross-correlation SNR, higher the amplitude and phase distortions.}
    \label{fig:lensing_pe}
\end{figure}

The residual cross-correlation informs us about the presence of a common feature in the residuals of both H1 and L1. To test whether the common feature in the residuals is present because of lensing, we estimate the lensing parameters from the data. We choose a physically-motivated lensing amplification model $A(f)$, defined as $A(f) = {\tilde{h}_{lensed}(f)} / {\tilde{h}_{unlensed}(f)}$, given by,

\begin{equation}\label{eq:amp}
    A(f) = a \left[1 + b \times e^{-kf} cos \left(\frac{2 \pi f}{f_0} + \phi_0 \right) \right] \exp \left(ib' \times e^{-k'f}cos \left(\frac{2 \pi f}{f_0} + \phi_0' \right) \right],
\end{equation}
where, for amplitude modulation, $a$ is the strong-lensing amplitude magnification $\equiv \sqrt{\mu}$ ; $b$ captures the perturbation scale at which microlensing affects the GW amplitude ; $f_0$ denotes the characteristic frequency of the oscillations in the amplification; $\phi_0$ denotes the phase-shift of the amplification and $e^{-kf}$ ensures that the amplification slowly converges to $a$ as the frequency $f$ is much higher than $1/k$. The phase modulation factor oscillates around zero with an amplitude $b'$ in the wave-optics regime and converges to zero as the frequency $f$ increases much more than $1/k'$. Note that this amplification model is only applicable if the lens mass distribution is spherically symmetric. For a performance comparison between this amplification template and the amplification from realistic astrophysical objects we refer to \citep{Chakraborty:2024mbr}. In the aforesaid work, we have compared Eq. \ref{eq:amp} to the wave-optics lensing amplification for point lens and singular isothermal sphere lens. We highlight that such an oscillatory trend in the amplification can be found for any generic lens and thus by tuning in the template structure and its parameters, it can represent arbitrary lens effects. However, this template does not account for the diffraction-limited regime of lensed GWs, where the amplification does not contain oscillations but becomes a monotonic function with frequency.

We estimate lensing parameters ($f_0, b, \phi_0, b', \phi_0'$) for the different waveform models using the Bayesian parameter estimation technique. Since the signal is observed in a very narrow frequency range, the frequency evolution terms ($e^{-kf}$ and $e^{-k'f}$) in the amplifications are dropped. These five parameters are sampled from a uniform prior, and the likelihood is Gaussian. A python-based parameter estimator tool \texttt{BILBY}\citep{Ashton:2018jfp} with the \texttt{DYNESTY}\citep{2020MNRAS.493.3132S} sampler is used. The sampling is performed with 500 live points and with a condition of stopping when the differential change of log(evidence) is 0.1. For more details on how these parameters are implemented in \texttt{DYNESTY}, see \citep{2020MNRAS.493.3132S}.

A wave-optics effect is observed if all of the following conditions hold:
\begin{enumerate}
    \item \textbf{Strong modulation of amplitude and phase:} In the lensing amplification model, $b$ captures the distortions of the GW amplitude due to lensing. Similarly, $b'$ captures the effect of GW phase distortions due to lensing. Non-zero estimations of these parameters point towards a feature which is caused by lensing. 

    \item \textbf{Intermediate values of the frequency scale:} In the lensing amplification model, $f_0$ represents the frequency scale with which amplitude and phase modulation vary. Whereas small $f_0$ can be the result of fitting the noise with fast oscillations, large $f_0$ values point towards a slow-to-no variation. An intermediate $f_0$, thus, can point towards a lensing feature. 
\end{enumerate}

In figure \ref{fig:lensing_pe}, we show the estimation of the lensing parameters for five waveform models. We present the marginalized single-parameter posteriors and the 2-dim joint contours using \texttt{CORNER} \citep{corner}. 
We observe that the amplitude-level lensing perturbation term $b$ for all waveforms agree to values close to zero except for \texttt{SEOBNRv5PHM}, which shows a support of up to $b=0.8$ at 95\% credible interval. This agrees with the fact that the residual cross-correlation SNR for this waveform model is the highest, at $7\sigma$. Also, for \texttt{IMRPhenomXPHM}, with residual SNR of $2.9$, we have support for $b$ up to 0.5. 
The phase-level lensing perturbation term $b'$ has the maximum support towards a value of zero, suggesting small to no amplitude and phase distortions of the signals. However, we note that \texttt{SEOBNRv5PHM} has a phase distortion support of up to 0.5, and \texttt{IMRPhenomTPHM} has a support for $b'$ of up to 0.7.
The estimation of the frequency scale $f_0$, shows no support for any particular frequency, and we obtain a uniform support for oscillations of all frequencies. Some support for the rapid oscillations is obtained for a few waveforms. However, since the oscillation amplitudes are small, these oscillations are fitting noise artifacts, without representing any physical process involved. 
The phase-shift estimates for amplitude modulation $\phi_0$ and phase modulation $\phi'_0$ for the different waveform models also tend to agree to some extent. However, with small $b$ and $b'$ and no support for moderately small-scale oscillations with $f_0$, these phase values do not contribute to distorting the waveform in a way similar to lensing in a conclusive way.

\section{Effects of waveform modeling uncertainties affecting the residual cross-correlation}\label{sec5}

The residual cross-correlation depends on the calculation of the best-fit waveform model. The waveform models are expected to agree well with one another up to the innermost stable circular orbit (ISCO) frequency, where post Newtonian approximations \citep{2014LRR....17....2B} are valid. For GW231123 like events we can calculate the ISCO frequency to be $\approx 15.4$Hz \footnote{We have calculated the ISCO frequency using the expression for Kerr black holes as found in \citep{1972ApJ...178..347B}. We have chosen the maximum likelihood masses and effective spin for the ISCO frequency calculation.}, which lies in the frequency ranges where detector noise is dominated by Earth's seismic activities. In fact, we have bandpassed anything below 20Hz, thus the observed residual is calculated for post-inspiral stages, where the agreement between waveform models is substantially worse and different waveform models start to move apart in amplitude and phase evolution of GWs. For high-mass GW events, in band durations are themselves short $\sim (\mathcal{O}(0.1s))$ and thus the characterization to the source is subjective to huge uncertainties since the waveform models can possess large systematic errors. Such waveform systematics can cause hindrance in the detections of both geometric-optics lensing \citep{Garron:2023gvd} and wave-optics lensing \citep{Mishra:2023ddt}.

To check whether a high residual cross-correlation SNR can arise due to uncertainties in waveform modeling, we calculated the mismatch between waveform models. The mismatch function between two waveforms is given by (maximized over time and phase),

\begin{equation}
    \mathscr{M}(h_1, h_2) = 1- \frac{\langle h_1| h_2\rangle}{\sqrt{\langle h_1| h_1\rangle \langle h_2| h_2\rangle}} \hspace{0.2cm},
\end{equation}
where the noise weighted inner product is defined as, 

\begin{equation}
    \langle a | b \rangle \equiv 4 \mathcal{R} \left[ \int_{f_{low}} ^{f_{high}} \frac{a(f) b^{*}(f)}{S_n(f)}df \right] \hspace{0.2cm}.
\end{equation}

Here, $S_n(f)$ is the power spectral density of the detector noise and $\mathcal{R}$ denotes the real part of the integral. 

We present the mismatch between the best-fit waveform models in the table \ref{tab:mismatch}. Compared to the \texttt{NRSur7dq4} model producing the lowest residual cross-correlation, we show that the mismatches of other waveforms vary in the range $4.2\%$ to $13.2\%$. The models producing residual cross-correlation SNR above the threshold of three show the highest mismatches with \texttt{NRSur7dq4} at $13.2\%$ (\texttt{IMRPhenomXO4a}, with $\rho_{residual}=4.7$) and $9.7\%$ (\texttt{SEOBNRv5PHM}, with $\rho_{residual}=7.2$). Compared to other waveforms, mismatches reach $24.7\%$ (between \texttt{IMRPhenomXO4a} and \texttt{IMRPhenomXPHM}) at maximum, implying a high degree of uncertainty between waveform models.

\begin{table}[h!]
\centering
\begin{tabular}{llllll}
\toprule
 & IMRPhenomXPHM & IMRPhenomTPHM & IMRPhenomXO4a & NRSur7dq4 & SEOBNRv5PHM \\
\midrule
IMRPhenomXPHM & 0.000 & 0.201 & 0.247 & 0.042 & 0.137 \\
IMRPhenomTPHM & - & 0.000 & 0.055 & 0.085 & 0.165 \\
IMRPhenomXO4a & - & - & 0.000 & 0.132 & 0.160 \\
NRSur7dq4 & - & - & - & 0.000 & 0.097 \\
SEOBNRv5PHM & - & - & - & - & 0.000 \\
\bottomrule
\end{tabular}
\caption{In the table, we have shown the mismatch between best-fit waveforms from different waveform models. Redundant entries have been skipped.}
\label{tab:mismatch}
\end{table}

\section{False lensing alarms for GW231123 arising from unlensed high-mass GW events}\label{sec6}

\begin{table}[h!]
\begin{tabular}{lccc}
\hline
$\mathcal{M}= 80 M_{\odot}, d_L=2600Mpc$\\ \hline
\hline
Parameter & IMRPhenomXPHM & IMRPhenomTPHM & IMRPhenomXO4a \\ \hline
$\mathcal{M} [M_{\odot}]$ & $81.27_{-2.41}^{+2.23}$ & $78.04_{-3.43}^{+2.49}$ & $79.79_{-2.20}^{+1.75}$ \\[6pt]
q & $0.89_{-0.10}^{+0.08}$ & $0.82_{-0.19}^{+0.12}$ & $0.85_{-0.10}^{+0.10}$ \\[6pt]
$a_1$ & $0.64_{-0.38}^{+0.19}$ & $0.56_{-0.25}^{+0.26}$ & $0.72_{-0.27}^{+0.17}$ \\[6pt]
$a_2$ & $0.24_{-0.18}^{+0.39}$ & $0.54_{-0.40}^{+0.34}$ & $0.42_{-0.29}^{+0.35}$ \\[6pt]
$\theta_1$ & $0.46_{-0.24}^{+0.44}$ & $0.45_{-0.25}^{+0.46}$ & $0.57_{-0.30}^{+0.45}$ \\[6pt]
$\theta_2$ & $1.07_{-0.63}^{+1.02}$ & $0.55_{-0.32}^{+0.81}$ & $1.11_{-0.59}^{+0.70}$ \\[6pt]
$d_L [Mpc]$ & $2172.07_{-379.30}^{+352.04}$ & $2257.96_{-295.98}^{+265.00}$ & $2299.35_{-380.98}^{+325.26}$ \\[6pt]
$\theta_{jn}$ & $0.56_{-0.23}^{+0.23}$ & $0.49_{-0.24}^{+0.27}$ & $0.50_{-0.23}^{+0.25}$ \\[6pt]
$\phi$ & $2.95_{-0.41}^{+2.88}$ & $4.38_{-2.56}^{+1.27}$ & $2.20_{-0.35}^{+2.59}$ \\[6pt]
\hline

\hline
$\mathcal{M}= 100 M_{\odot}, d_L=3000Mpc$ \\ \hline
\hline
Parameter & IMRPhenomXPHM & IMRPhenomTPHM & IMRPhenomXO4a \\ \hline
$\mathcal{M} [M_{\odot}]$ & $104.97_{-3.29}^{+2.91}$ & $95.75_{-3.73}^{+3.01}$ & $100.66_{-3.50}^{+2.64}$ \\[6pt]
q & $0.76_{-0.13}^{+0.13}$ & $0.79_{-0.21}^{+0.14}$ & $0.83_{-0.14}^{+0.11}$ \\[6pt]
$a_1$ & $0.80_{-0.20}^{+0.12}$ & $0.58_{-0.25}^{+0.23}$ & $0.69_{-0.20}^{+0.19}$ \\[6pt]
$a_2$ & $0.18_{-0.14}^{+0.31}$ & $0.45_{-0.35}^{+0.39}$ & $0.70_{-0.32}^{+0.21}$ \\[6pt]
$\theta_1$ & $0.29_{-0.15}^{+0.24}$ & $0.38_{-0.21}^{+0.46}$ & $0.73_{-0.32}^{+0.37}$ \\[6pt]
$\theta_2$ & $1.07_{-0.66}^{+0.90}$ & $0.65_{-0.40}^{+0.97}$ & $0.91_{-0.39}^{+0.45}$ \\[6pt]
$d_L [Mpc]$ & $2907.44_{-389.15}^{+330.63}$ & $2463.26_{-352.00}^{+307.45}$ & $2348.39_{-447.84}^{+455.05}$ \\[6pt]
$\theta_{jn}$ & $0.38_{-0.17}^{+0.19}$ & $0.54_{-0.27}^{+0.27}$ & $0.61_{-0.22}^{+0.22}$ \\[6pt]
$\phi$ & $1.85_{-0.42}^{+0.41}$ & $3.98_{-2.79}^{+1.68}$ & $1.68_{-0.40}^{+0.46}$ \\[6pt]
\hline

\hline
$\mathcal{M}= 120 M_{\odot}, d_L=3150Mpc$\\ \hline
\hline
Parameter & IMRPhenomXPHM & IMRPhenomTPHM & IMRPhenomXO4a \\ \hline
$\mathcal{M} [M_{\odot}]$ & $111.96_{-6.64}^{+5.28}$ & $107.02_{-5.50}^{+4.02}$ & $111.02_{-5.08}^{+4.05}$ \\[6pt]
q & $0.78_{-0.16}^{+0.14}$ & $0.71_{-0.14}^{+0.15}$ & $0.71_{-0.16}^{+0.16}$ \\[6pt]
$a_1$ & $0.39_{-0.26}^{+0.30}$ & $0.17_{-0.12}^{+0.22}$ & $0.50_{-0.30}^{+0.25}$ \\[6pt]
$a_2$ & $0.34_{-0.25}^{+0.33}$ & $0.52_{-0.34}^{+0.31}$ & $0.50_{-0.32}^{+0.31}$ \\[6pt]
$\theta_1$ & $1.21_{-0.64}^{+0.63}$ & $1.16_{-0.63}^{+0.95}$ & $1.13_{-0.49}^{+0.50}$ \\[6pt]
$\theta_2$ & $1.35_{-0.72}^{+0.75}$ & $0.92_{-0.47}^{+0.69}$ & $1.39_{-0.63}^{+0.68}$ \\[6pt]
$d_L [Mpc]$ & $2479.17_{-352.31}^{+388.73}$ & $2206.15_{-270.62}^{+292.26}$ & $2649.92_{-340.64}^{+339.36}$ \\[6pt]
$\theta_{jn}$ & $0.50_{-0.21}^{+0.21}$ & $0.46_{-0.18}^{+0.23}$ & $0.42_{-0.17}^{+0.19}$ \\[6pt]
$\phi$ & $2.63_{-0.39}^{+0.58}$ & $2.45_{-1.97}^{+3.25}$ & $2.27_{-0.39}^{+0.46}$ \\[6pt]
\hline
\end{tabular}
\caption{The table shows the estimated source properties for a set of three different events with similar intrinsic source properties to that of GW231123. The events are of similar SNR to GW231123, with source chirp-masses (in the detector frame) being 80, 100, 120 $M_{\odot}$ respectively. The error-bars represent the $\pm 1\sigma$ uncertainties around the median associated with the inference. The signal is generated with the waveform model \texttt{IMRPhenomXPHM-SpinTaylor} and the inferences are performed with \texttt{IMRPhenomXPHM-SpinTaylor}, \texttt{IMRPhenomTPHM} and \texttt{IMRPhenomXO4a}.}
\label{Table:tab1}
\end{table}

One of the potential reasons for the study of GW231123 is the inferred high masses of its BBH source, which increases the possibility of a lensed GW signal. To understand whether waveform systematic error alone for a short event like GW231123 can significantly move the estimation of inferred source properties away from the true properties, we simulated unlensed GW events with very high source masses similar to that of GW231123, and estimated their source parameters to observe whether such events can pose as false alarms in lensing \citep{Read:2023hkv, Garron:2023gvd, Keitel:2024brp}. We performed this test to observe the degree of posterior disagreement as supported by the waveform uncertainties, feasible by high-mass unlensed events, and to observe how that compares to GW231123 posteriors for different waveform models. This helps us to understand how any meaningful residual cross-correlation SNR can make a source appear as lensed because of the presence of considerable systematic error, even when the source is not lensed.

We have simulated three GW events from BBH mergers with chirp masses ($\mathcal{M}$) $80M_{\odot}$, $100M_{\odot}$, $120M_{\odot}$ with a mass ratio ($q$) $0.75$. The spin magnitudes of the components ($a_1$ and $a_2$) are kept at 0.5, with their spins aligned along the direction of the orbital angular momentum ($\theta_1 = \theta_2 = 0$). The inclination angle ($\theta_{jn}$) and the coalescence phase ($\phi$) are chosen as 0.4 and 1.3 respectively. The luminosity distance ($d_L$) of these events is chosen in such a way that the matched-filter SNR across the H1-L1 detector network is $\approx$ 22.8, almost equal to that of GW231123. The distances to the GW sources of the aforementioned chirp masses are 2600Mpc, 3000Mpc, and 3150Mpc, respectively, to adjust the SNR of the event to lie close to 22.8. These three events were generated with the \texttt{IMRPhenomXPHM-SpinTaylor} \citep{Pratten:2020ceb} waveform model. The source parameters are recovered with three different waveform models: \texttt{IMRPhenomXPHM-SpinTaylor}\citep{Pratten:2020ceb}, \texttt{IMRPhenomTPHM}\citep{Estelles:2021gvs}, and \texttt{IMRPhenomXO4a}\citep{Thompson:2023ase}. The recovered parameter distributions for the chirp mass $80M_{\odot}$, $100M_{\odot}$, $120M_{\odot}$ are shown in the figures \ref{fig:fig_pe_80_far}, \ref{fig:fig_pe_100_far} and \ref{fig:fig_pe_120_far} in the appendix \ref{app:2}. The same information of these corner-plots is presented in the table \ref{Table:tab1}.

We show that as the GW source masses increase, the estimates of the masses start to deviate more from the true value, even when the same matched-filter SNR is kept the same with the same noise. At 80$M_{\odot}$, the chirp mass-median is off by $\leq 1\sigma$ for all three waveforms. However, at 100$M_{\odot}$ the chirp mass-median is off by $1\sigma-2\sigma$. At even higher chirp masses 120$M_{\odot}$ the chirp mass-median is off by $2\sigma-3\sigma$. The same trend is followed across the considered waveform models and, it occurs even when the parameter recovery is performed with the same waveform model as the data generation. These results describe that even without lensing, the high level of waveform systematics associated with GWs from heavy BBH merger events. Such waveform systematic error can, in turn, produce correlated residuals that can show up in cross-correlation and mimic features that resemble lensing. A similar test of spin-dependence of waveform systematics that may affect a microlensing detection is performed in the appendix \ref{app:3}.

\section{Residual Tests on GWTC-4 events with cross-correlation}\label{gwtc4all}

We examined the events in the GWTC-4 catalog \citep{LIGOScientific:2025slb, LIGOScientific:2025pvj} to search for any wave-optics lensing signatures using the residual cross-correlation technique. We obtain 76 GW events with a matched-filter network SNR above 8 and observed with at least two detectors. To have a robust detection method, we have considered the part of the residual up to the frequency $f_{\rm isco}$ of the BBH inspiral phase, up to which the systematic error in waveform modeling is significantly lower than the following merger-ringdown phases. The procedure for detecting modulations due to lensing, followed here, is the same as in our previous work \citep{Chakraborty:2025maj}, where we applied the cross-correlation technique in the GWTC-3 catalog. Using the best-fit GW waveform using the \texttt{IMRPhenomXPHM-SpinTaylor} model, we obtain the residuals and calculate residual SNR by cross-correlating the residuals and summing the signal between the timestamps $t_{20}$ and $t_{\rm isco}$ \footnote{$t_{20}$ and $t_{\rm isco}$ correspond to the time when the GW emission is predominant at 20Hz and $f_{\rm isco}$ respectively.}. 

\begin{figure}
    \centering
    \includegraphics[width=\linewidth]{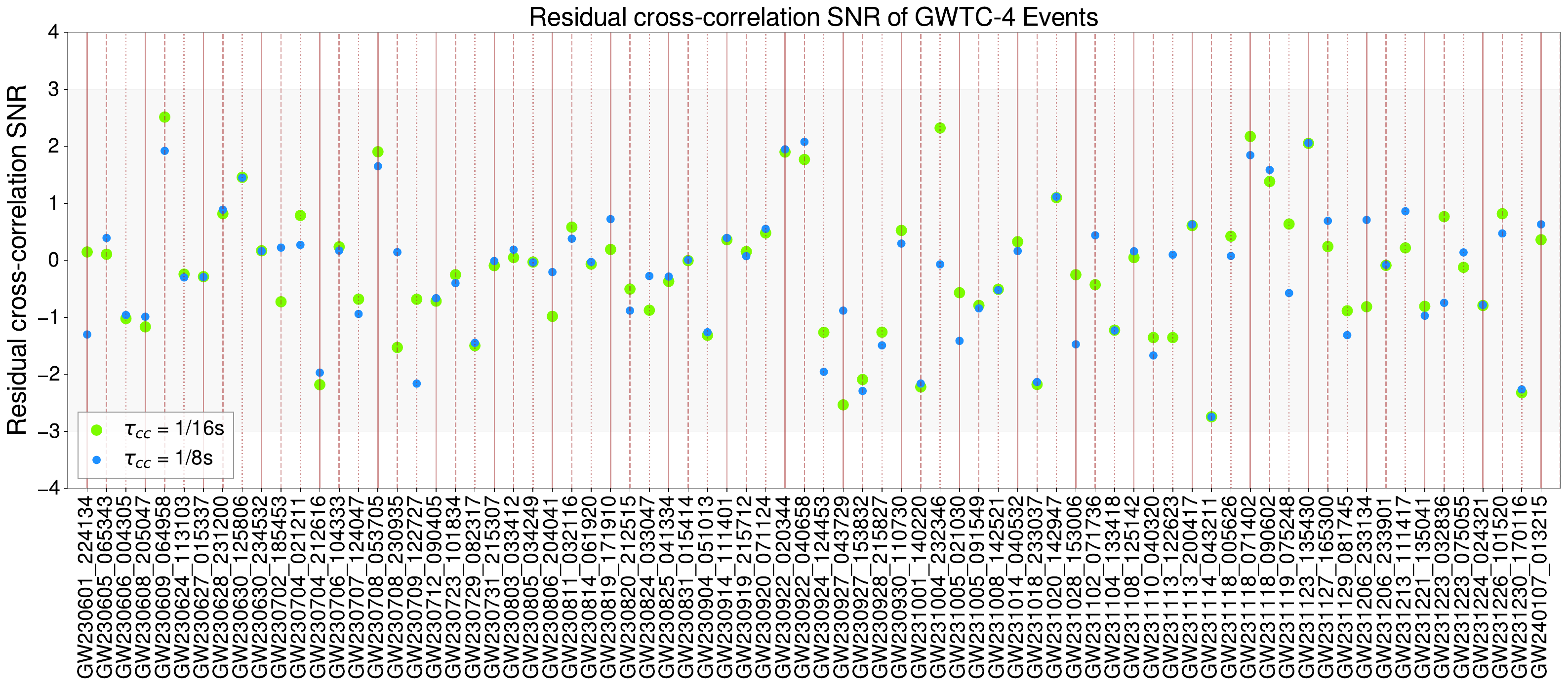}
    \caption{In this figure, we present our findings from the searches of wave-optics lensing features from the GWTC-4 catalog. The horizontal axis shows the events in chronological sequence, and the vertical axis shows the residual SNR. To keep waveform modeling errors in check, we considered only the portion of the waveform when the BBH orbit is larger than the innermost stable circular orbit (ISCO) the frequency of the GW is lower than $f_{\rm isco}$.}
    \label{fig:gwtc4}
\end{figure}

In figure \ref{fig:gwtc4}, we present the key results from our wave-optics searches of the GWTC-4 events. The residual SNRs have been calculated for the cross-correlation timescales of $1/16s$ and $1/8s$. The significance of GW231123 decreases from $\approx 2.9\sigma$ to $\approx 2.0\sigma$ as we remove the post-inspiral phases from this analysis. We find no evidence of a wave-optics lensing modulation from any of the events with a statistical significance of or above 3$\sigma$. The maximum significance is obtained for the event GW231114\_043211 with a 2.75$\sigma$ deviation from the noise statistics. 

\section{Conclusions and Future Prospects} \label{sec7}

In this work, we search for lensing signatures in the event GW231123 using model-independent residual tests. We perform cross-correlation between the residuals of the detectors to check the strength of their common features as compared to the noise. We perform a Bayesian analysis to characterize the frequency-dependent amplitude and phase modulations. Both tests lead to little-to-no support for the evidence of lensing from the event, with the lowest support for wave-optics lensing coming from the surrogate numerical relativity waveform model \texttt{NRSur7dq4} at $1.4\sigma$. Thus, we do not detect any wave-optics lensing features from GW231123. However, the waveform systematics effect for this kind of high-mass events shows considerable waveform discrepancies at both amplitude and phase level. We show that even the source parameter estimates from an unlensed GW231123-like event can be largely off, even with detection at high matched-filter SNR ($\sim 22.8$). The mismatch between the best-fit waveforms from different models is observed to be up to $\sim 25\%$. The current set of waveform models is, therefore, not compatible with the characterization of a GW source in the high-mass regime. 

However, if the event GW231123 is lensed, one would expect to detect similar events in the future, which can confirm the lensing hypothesis. Given that we observe one lensed event from events up to all four observational runs (O1+O2+O3+O4a), we calculate the number of detectable lensed events from future LVK runs. We find that similarly amplified lensed events are expected to be observed at $\approx 4.08 \pm 2.02$ with O4-like detector sensitivities of H1 and L1 for a total runtime of 3 years with a duty factor of 0.75. A higher magnified GW signal than GW231123 is rarer and a lower magnification is more frequent due to the impact of the lensing optical depth \citep{Mukherjee:2020tvr}. Thus, the observation of the aforementioned number of lensed events in the given time-frame would allow for a more confident claim of the lensing hypothesis of GW231123.

There can be unaccounted physical effects associated with a BBH merger that, if not considered during modeling, can introduce lensing-like signatures. Thus, it is possible that the waveform systematic error can introduce effects similar to wave-optics lensing features. However, in the current stage, we show that inference of masses from an unlensed high-mass event can show deviations when different waveforms are implemented to recover source parameters. Therefore, accurate waveforms for high-mass systems such as GW231123 are required first to constrain the source properties, before introducing any lensing hypothesis. A detector to observe a lower-band in the detectable frequencies like deciHz like TianGo \citep{Kuns:2019upi} or mHz like LISA \citep{Babak:2021mhe}, would allow for better constraints of the properties of intermediate-mass BBH GW sources like GW231123. Due to significantly longer timescales of observation than in the LVK-band, waveform modeling associated with these events can proceed in a data-driven way. This would compensate for the loss of the signal in the kHz-band, where IMBH detections are rare and thus waveforms are not built specifically for those events, resulting in waveforms with high systematics. 

Improving the sensitivity of the LVK detectors \citep{KAGRA:2013rdx} would allow better constraints on the properties of the merging BBH. With the upcoming GW observatories like LIGO-India \citep{LIGO_India}, Cosmic-Explorer \citep{Punturo:2010zza, Reitze:2019iox}, Einstein Telescope \citep{Hild:2010id, LIGOScientific:2016wof}, more detections of GW events can help us to understand the theoretical PISN mass gap, if this actually exists. The high systematic error associated with the modeling of a high-mass event such as GW231123, although its significant detection SNR and masses in the PISN mass gap, does not provide conclusive evidence of the lensing of the event.

\begin{acknowledgments}
The authors express their gratitude to the  members of the \texttt{⟨data|theory⟩ Universe-Lab} group for useful suggestions. This work is part of the \texttt{⟨data|theory⟩ Universe-Lab}, supported by TIFR and the Department of Atomic Energy, Government of India. The authors express gratitude to the computer cluster of \texttt{⟨data|theory⟩ Universe-Lab} for computing resources used in this analysis. We thank the LIGO-Virgo-KAGRA Collaboration for providing noise curves. LIGO, funded by the U.S. National Science Foundation (NSF), and Virgo, supported by the French CNRS, Italian INFN, and Dutch Nikhef, along with contributions from Polish and Hungarian institutes. The research leverages data and software from the Gravitational Wave Open Science Center, a service provided by LIGO Laboratory, the LIGO Scientific Collaboration, Virgo Collaboration, and KAGRA. Advanced LIGO's construction and operation receive support from STFC of the UK, Max-Planck Society (MPS), and the State of Niedersachsen/Germany, with additional backing from the Australian Research Council. Virgo, affiliated with the European Gravitational Observatory (EGO), secures funding through contributions from various European institutions. Meanwhile, KAGRA's construction and operation are funded by MEXT, JSPS, NRF, MSIT, AS, and MoST. This material is based upon work supported by NSF’s LIGO Laboratory which is a major facility fully funded by the National Science Foundation. We acknowledge the use of the following python packages in this work: NUMPY \citep{harris2020array}, SCIPY \citep{2020SciPy-NMeth}, MATPLOTLIB \citep{Hunter:2007}, PYCBC \citep{alex_nitz_2024_10473621}, GWPY \citep{gwpy}, LALSUITE \citep{lalsuite}, BILBY \citep{Ashton_2019}, PESUMMARY \citep{Hoy:2020vys}, CORNER \citep{corner}.

\end{acknowledgments}

\bibliography{biblio}{}
\bibliographystyle{aasjournalv7}

\appendix

\section{Brief description of $\rm \mu$-\texttt{GLANCE}}\label{app:1}

To look for any signature beyond the one modeled in a GW signal, we have developed the residual cross-correlation-based technique $\mu$-\texttt{GLANCE}\citep{Chakraborty:2024mbr}, which does not assume any model for the lens structure. A truly microlensing feature in the GW must be common to all detectors. $\mu$-\texttt{GLANCE} uses this idea to find the common features present in the residuals of different detectors while suppressing uncorrelated detector noise. 

Given the posterior distributions of all source intrinsic and extrinsic parameters of a GW event, we can construct the best-fit GW waveform (corresponding to the maximum-likelihood values of the source parameters). The best-fit GW waveform projected on to each detector is subtracted from the data to obtain the residual: $R_i(t) \equiv d_i(t) - h_{ BF,  i}(t)$, here $R_i(t)$ is the residual at the detector $i$, calculated by subtraction of the detector specific best-fit waveform $ h_{BF, i}$ from the data $d_{i}(t)$. If a microlensing signature is present in the GW signal,  which is not captured by the best-fit template, it must be present in the residuals from all detectors. Therefore, cross-correlating the residuals from different detectors, we can detect for any beyond modeled signature. The cross-correlation of the residuals between detectors $x$ and $x'$ is given by,
\begin{equation}\label{eq2}
    D_{x x'} (t) \equiv R_x \otimes R_{x'} = \frac{1}{\tau_{cc}} \int_{t-\tau_{cc}/2} ^{t+\tau_{cc}/2} R_{x} (t') R_{x'} (t' + t_d) dt',   
\end{equation}
where, $R_i(t) = r_i(t) + n_i(t)$, here $r_i(t)$ denotes the residual GW, $n_i(t)$ denotes the detector noise, and $t_d$ incorporates the time-delay between the arrival of the signal at different detectors due to the finite propagation speed of GWs. The cross-correlation timescale ($\tau_{cc}$) is typically chosen to be of the order or shorter than the signal duration.

A truly microlensing feature is spread across the GW frequency range and thus is present at all times during the signal. A genuine lensing feature is therefore spread across the entirety of the signal and is non-localized. In contrast, any effect from waveform systematics, detector glitch causes a sudden change in the cross-correlation. To understand the non-localized nature of the residual cross-correlation and to observe how the residual cross-correlation builds up over time, we sum over the time-domain residual cross-correlation values between $[t_0, t_0 + \tau_{snr}]$\footnote{Here, $t_0$ and $\tau_{snr}$ are chosen to capture the possible effects in the time-domain signal completely.}, when the signal is present, given by,
\begin{equation}\label{eq:sum_cc}
    S_{xx'} =  \sum _ {t'=t_0} ^{t'= t_0 +\tau_{snr}} D_{xx'}(t')  \hspace{0.25cm}.
\end{equation}
In order to quantify the statistical significance of the  residual cross-correlation signal $S_{xx'}$, we compare it with the standard deviation of the time-summed noise cross-correlation $N_{xx'}$ defined as,

\begin{equation}
    N_{x x'} (t) \equiv n_x \otimes n_{x'} = \frac{1}{\tau_{cc}} \int_{t-\tau_{cc}/2} ^{t+\tau_{cc}/2} n_{x} (t') n_{x'} (t') dt'.
\end{equation}

\begin{figure}
    \centering
    \includegraphics[width=\linewidth]{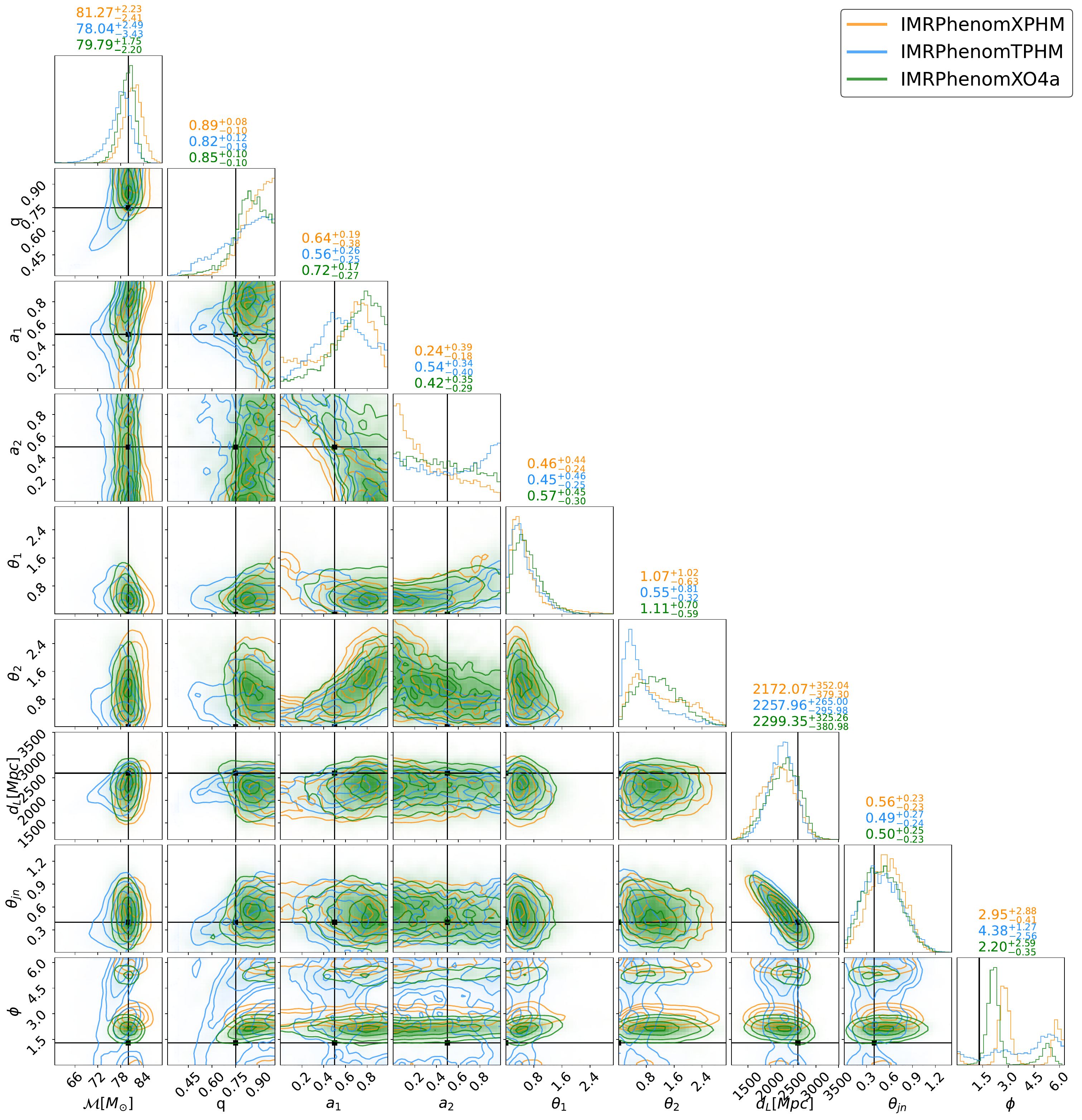}
    \caption{In the figure, we present the recovery of different GW source parameters by a set of waveform models for a simulated heavy-mass ($\mathcal{M}_c = 80 M_{\odot}$, $q=0.75$) signal with \texttt{IMRPhenomXPHM-SpinTaylor} waveform with SNR similar to GW231123 ($\rho_N = 22.78$). The recovery has been performed with a set of three different waveform models: \texttt{IMRPhenomTPHM}, \texttt{IMRPhenomXO4a} and \texttt{IMRPhenomXPHM-SpinTaylor}.}
    \label{fig:fig_pe_80_far}
\end{figure}

\begin{figure}
    \centering
    \includegraphics[width=\linewidth]{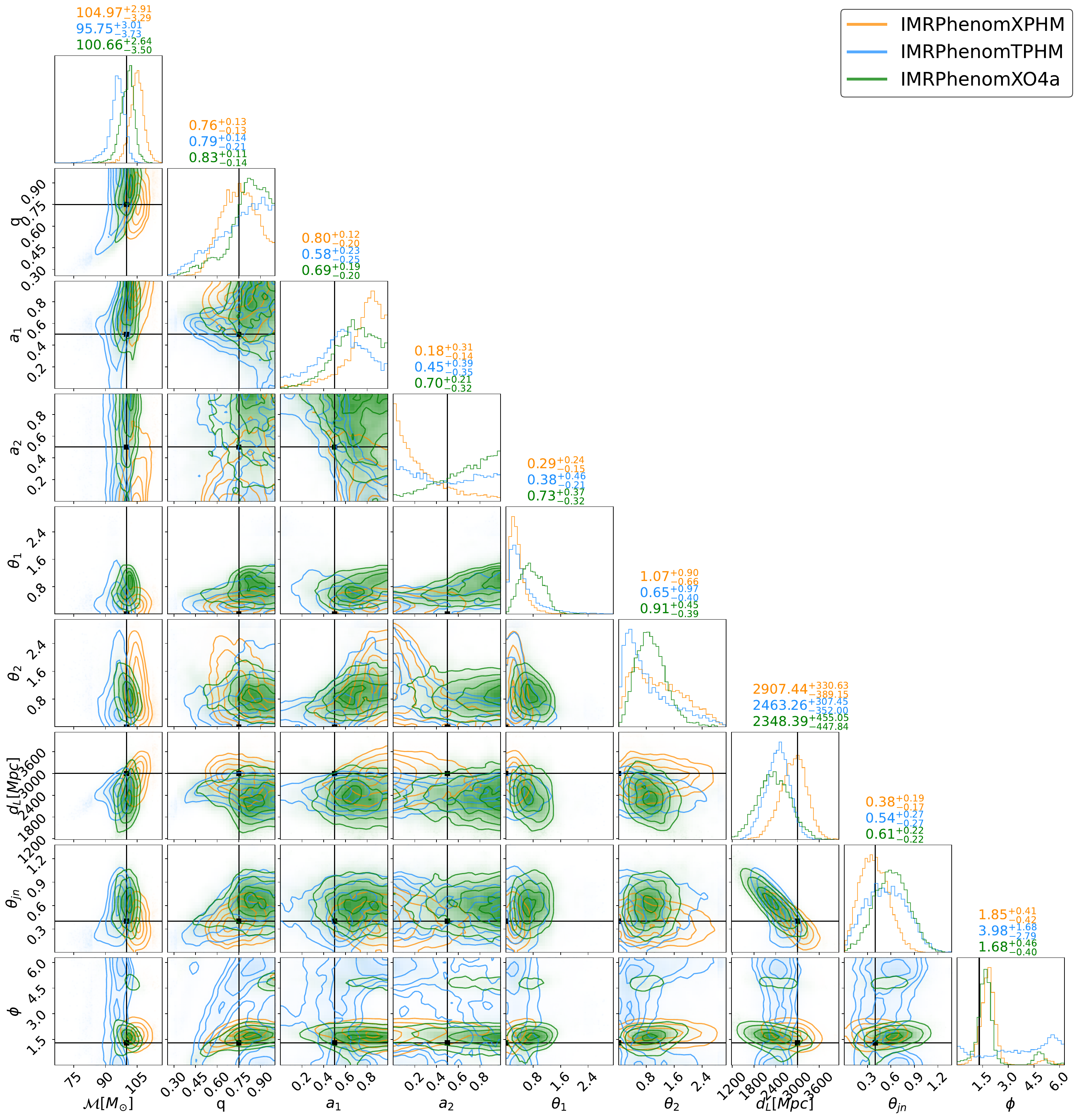}
    \caption{In the figure, we present the recovery of different GW source parameters by a few different waveforms for a simulated heavy-mass ($\mathcal{M}_c = 100 M_{\odot}$, $q=0.75$) signal with \texttt{IMRPhenomXPHM-SpinTaylor} waveform with SNR similar to GW231123 ($\rho_N = 22.77$). The recovery has been performed with a set of three different waveform models: \texttt{IMRPhenomTPHM}, \texttt{IMRPhenomXO4a} and \texttt{IMRPhenomXPHM-SpinTaylor}.}
    \label{fig:fig_pe_100_far}
\end{figure}

We take the detector noise from the data before and after the signal was present. We calculate the time-stacked noise cross-correlation for the same $\tau_{\rm snr}$. We estimate the standard deviation (denoted by $\sigma_{xx'}$) of these stacked noise cross-correlation using the relation,
\begin{equation}
    \sigma_{xx'} = \sqrt{ \textbf{Var} \left(\sum_{t'=t} ^{t'=t+\tau_{\rm snr}} N_{xx'} (t')\right) } \hspace{0.25cm},
\end{equation}
here we denote the variance of the noise cross-correlation residual by \textbf{Var}.
Using the signal cross-correlation and noise cross-correlation defined above, we define the residual cross-correlation signal-to-noise ratio (SNR) as,
\begin{equation}\label{eq:snr}
    \rho_{\rm residual} \equiv \frac{S_{xx'}}{\sigma_{xx'}} = \frac{  \sum _ {t_0} ^{t_0 + \tau_{snr}} D_{xx'}}{\sqrt{ \textbf{Var} \left(\sum_{t} ^{t + \tau_{\rm snr}} N_{xx'} \right) }} \hspace{0.25cm}.
\end{equation}

The quantity $\rho_{\rm residual}$ captures the deviation of the signal strength compared to the strength of the noise cross-correlation. We have performed the first-model-independent search for microlensing from LVK data up to GWTC-3 \citep{Chakraborty:2025maj} using \texttt{$\mu$-GLANCE}. As of the current state, no microlensing evidence has been found towards any event. An approach to look for microlensing signatures from strongly lensed GWs can be found here \citep{Seo:2025dto}.

\section{Posterior distributions for high-mass events using different waveform models}\label{app:2}

\begin{figure}
    \centering
    \includegraphics[width=\linewidth]{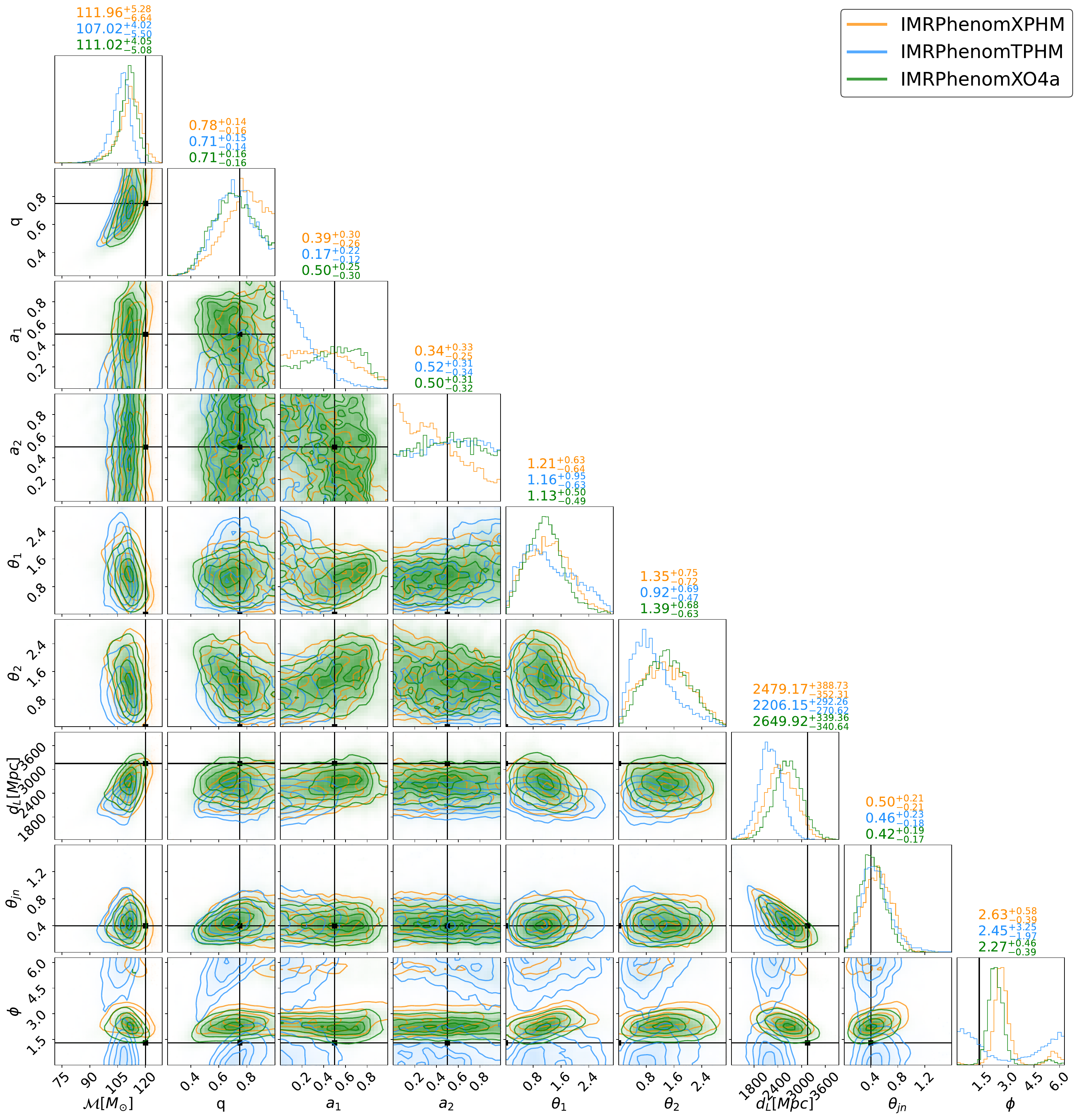}
    \caption{In the figure, we present the recovery of different GW source parameters by a few different waveforms for a simulated heavy-mass ($\mathcal{M}_c = 120 M_{\odot}$, $q=0.75$) signal with \texttt{IMRPhenomXPHM-SpinTaylor} waveform with SNR similar to GW231123 ($\rho_N = 22.80$). The recovery has been performed with a set of three different waveform models: \texttt{IMRPhenomTPHM}, \texttt{IMRPhenomXO4a} and \texttt{IMRPhenomXPHM-SpinTaylor}.}
    \label{fig:fig_pe_120_far}
\end{figure}

In this appendix, we present the marginalized posterior distributions of the GW source properties in 1-dimension and 2-dimensions using \texttt{CORNER}. In figures \ref{fig:fig_pe_80_far}, \ref{fig:fig_pe_100_far} and \ref{fig:fig_pe_120_far}, we show the inference of the GW source parameters, injected with a source chirp mass of 80$M_{\odot}$, 100$M_{\odot}$, and 120$M_{\odot}$, respectively. Three different waveform models used to recover the source properties are mentioned in the label in the top right corner. The figures show how waveform differences at the modeling level can introduce differences in the inference of the masses and spins of the merging BBH.

\section{Checking the spin-systematics of high-mass BBH mergers in explaining the discrepancies found in the posterior distribution of GW231123}\label{app:3}

\begin{table}
\centering
\setlength{\tabcolsep}{15pt}
\begin{tabular}{lllll}

\toprule
Parameter & Truth & IMRPhenomXPHM & IMRPhenomTPHM & IMRPhenomXO4a \\
\midrule
$\mathcal{M}_c [M_{\odot}]$ & $100.00$ & $99.903^{+3.398}_{-4.022}$ & $102.311^{+3.736}_{-3.515}$ & $100.007^{+3.052}_{-4.717}$ \\
q & $0.75$ & $0.843^{+0.109}_{-0.147}$ & $0.857^{+0.098}_{-0.146}$ & $0.828^{+0.116}_{-0.165}$ \\
$a_1$ & $0.10$ & $0.235^{+0.264}_{-0.169}$ & $0.427^{+0.338}_{-0.285}$ & $0.310^{+0.319}_{-0.222}$ \\
$a_2$ & $0.10$ & $0.356^{+0.384}_{-0.260}$ & $0.470^{+0.319}_{-0.327}$ & $0.497^{+0.319}_{-0.359}$ \\
$\theta_1$ & $0.00$ & $1.470^{+0.795}_{-0.772}$ & $1.459^{+0.614}_{-0.539}$ & $1.510^{+0.649}_{-0.679}$ \\
$\theta_2$ & $0.00$ & $2.127^{+0.500}_{-0.852}$ & $1.531^{+0.652}_{-0.644}$ & $2.076^{+0.545}_{-0.658}$ \\
$d_L [Mpc]$ & $3000.00$ & $2846.450^{+288.100}_{-334.857}$ & $2678.537^{+311.537}_{-330.776}$ & $2847.693^{+345.708}_{-358.605}$ \\
$\theta_{jn}$ & $0.40$ & $0.364^{+0.223}_{-0.187}$ & $0.324^{+0.264}_{-0.173}$ & $0.386^{+0.228}_{-0.183}$ \\
$\phi$ & $1.30$ & $4.974^{+0.800}_{-2.636}$ & $2.913^{+1.913}_{-1.745}$ & $5.213^{+0.585}_{-2.915}$ \\
\bottomrule

\toprule
Parameter & Truth & IMRPhenomXPHM & IMRPhenomTPHM & IMRPhenomXO4a \\
\midrule
$\mathcal{M}_c [M_{\odot}]$ & $100.00$ & $101.708^{+4.467}_{-4.566}$ & $101.713^{+2.831}_{-3.126}$ & $100.502^{+3.819}_{-3.730}$ \\
q & $0.75$ & $0.861^{+0.099}_{-0.126}$ & $0.889^{+0.075}_{-0.107}$ & $0.807^{+0.135}_{-0.145}$ \\
$a_1$ & $0.25$ & $0.416^{+0.307}_{-0.296}$ & $0.513^{+0.288}_{-0.337}$ & $0.639^{+0.241}_{-0.384}$ \\
$a_2$ & $0.25$ & $0.263^{+0.342}_{-0.189}$ & $0.422^{+0.361}_{-0.288}$ & $0.579^{+0.299}_{-0.350}$ \\
$\theta_1$ & $0.00$ & $0.754^{+0.808}_{-0.438}$ & $0.832^{+0.735}_{-0.407}$ & $0.999^{+0.546}_{-0.462}$ \\
$\theta_2$ & $0.00$ & $1.504^{+0.845}_{-0.843}$ & $1.262^{+0.813}_{-0.624}$ & $1.798^{+0.583}_{-0.801}$ \\
$d_L [Mpc]$ & $3000.00$ & $2753.390^{+469.418}_{-484.566}$ & $2303.232^{+549.225}_{-490.644}$ & $2878.542^{+404.781}_{-446.180}$ \\
$\theta_{jn}$ & $0.40$ & $0.503^{+0.243}_{-0.216}$ & $0.734^{+0.359}_{-0.335}$ & $0.416^{+0.206}_{-0.200}$ \\
$\phi$ & $1.30$ & $1.029^{+3.033}_{-0.646}$ & $2.456^{+2.138}_{-1.393}$ & $0.892^{+2.938}_{-0.546}$ \\
\bottomrule

\toprule
Parameter & Truth & IMRPhenomXPHM & IMRPhenomTPHM & IMRPhenomXO4a \\
\midrule
$\mathcal{M}_c [M_{\odot}]$ & $100.00$ & $105.077^{+3.412}_{-3.181}$ & $99.902^{+3.110}_{-2.716}$ & $102.599^{+2.491}_{-2.982}$ \\
q & $0.75$ & $0.806^{+0.119}_{-0.155}$ & $0.790^{+0.107}_{-0.106}$ & $0.738^{+0.144}_{-0.116}$ \\
$a_1$ & $0.50$ & $0.791^{+0.127}_{-0.186}$ & $0.662^{+0.181}_{-0.206}$ & $0.834^{+0.106}_{-0.180}$ \\
$a_2$ & $0.30$ & $0.287^{+0.354}_{-0.220}$ & $0.344^{+0.331}_{-0.243}$ & $0.590^{+0.288}_{-0.379}$ \\
$\theta_1$ & $0.00$ & $0.335^{+0.232}_{-0.174}$ & $0.483^{+0.393}_{-0.275}$ & $0.655^{+0.380}_{-0.302}$ \\
$\theta_2$ & $0.00$ & $1.119^{+0.887}_{-0.659}$ & $0.952^{+0.824}_{-0.526}$ & $1.394^{+0.554}_{-0.603}$ \\
$d_L [Mpc]$ & $3000.00$ & $2969.933^{+365.882}_{-437.589}$ & $2274.364^{+403.805}_{-394.586}$ & $2916.915^{+453.424}_{-517.021}$ \\
$\theta_{jn}$ & $0.40$ & $0.423^{+0.221}_{-0.195}$ & $0.784^{+0.229}_{-0.269}$ & $0.407^{+0.217}_{-0.196}$ \\
$\phi$ & $1.30$ & $2.208^{+0.538}_{-0.442}$ & $1.865^{+1.686}_{-1.066}$ & $2.196^{+0.461}_{-0.443}$ \\
\bottomrule

\toprule
Parameter & Truth & IMRPhenomXPHM & IMRPhenomTPHM & IMRPhenomXO4a \\
\midrule
$\mathcal{M}_c [M_{\odot}]$ & $100.00$ & $101.925^{+2.698}_{-3.069}$ & $98.244^{+2.766}_{-2.916}$ & $98.832^{+2.668}_{-2.149}$ \\
q & $0.75$ & $0.890^{+0.074}_{-0.113}$ & $0.796^{+0.120}_{-0.163}$ & $0.911^{+0.065}_{-0.125}$ \\
$a_1$ & $0.50$ & $0.721^{+0.181}_{-0.313}$ & $0.627^{+0.199}_{-0.244}$ & $0.698^{+0.193}_{-0.295}$ \\
$a_2$ & $0.50$ & $0.253^{+0.334}_{-0.189}$ & $0.490^{+0.334}_{-0.340}$ & $0.457^{+0.327}_{-0.303}$ \\
$\theta_1$ & $0.00$ & $0.412^{+0.335}_{-0.222}$ & $0.537^{+0.366}_{-0.266}$ & $0.808^{+0.518}_{-0.453}$ \\
$\theta_2$ & $0.00$ & $1.096^{+0.918}_{-0.646}$ & $0.833^{+0.781}_{-0.422}$ & $1.323^{+0.679}_{-0.607}$ \\
$d_L [Mpc]$ & $3000.00$ & $2562.606^{+437.504}_{-460.003}$ & $2546.619^{+297.360}_{-401.031}$ & $2339.045^{+482.585}_{-459.328}$ \\
$\theta_{jn}$ & $0.40$ & $0.568^{+0.224}_{-0.239}$ & $0.527^{+0.322}_{-0.265}$ & $0.563^{+0.218}_{-0.229}$ \\
$\phi$ & $1.30$ & $3.545^{+0.466}_{-2.714}$ & $2.057^{+1.413}_{-0.933}$ & $3.702^{+0.462}_{-2.766}$ \\
\bottomrule

\end{tabular}
\caption{In this table, we show the waveform systematics effects associated to the merger of heavy mass BBHs, by showing the deviations in the estimations of the source parameters in low spin regime. We inject the signal with \texttt{NRSur7dq4} waveform-model and recover the source properties using \texttt{IMRPhenomXPHM-SpinTaylor}, \texttt{IMRPhenomTPHM} and \texttt{IMRPhenomXO4a} models. All four parts of the table capture the systematics in the high spin regime $a_1, a_2 \leq 0.5$, with aligned spins. The distribution shapes and the correlation between different source parameters can be found in figure \ref{fig:spin_0.1} ($a_1 =a_2 = 0.1$ with aligned spins).}
\label{tab:spin_posteriors_low_spin}
\end{table}

\begin{table}
\centering
\setlength{\tabcolsep}{15pt}
\begin{tabular}{lllll}
\toprule
Parameter & Truth & IMRPhenomXPHM & IMRPhenomTPHM & IMRPhenomXO4a \\
\midrule
$\mathcal{M}_c [M_{\odot}]$ & $100.00$ & $104.870^{+2.921}_{-3.126}$ & $102.254^{+2.989}_{-3.767}$ & $102.076^{+2.767}_{-3.023}$ \\
q & $0.75$ & $0.698^{+0.139}_{-0.107}$ & $0.765^{+0.135}_{-0.154}$ & $0.662^{+0.090}_{-0.082}$ \\
$a_1$ & $0.75$ & $0.876^{+0.081}_{-0.116}$ & $0.721^{+0.165}_{-0.143}$ & $0.885^{+0.070}_{-0.099}$ \\
$a_2$ & $0.25$ & $0.502^{+0.255}_{-0.274}$ & $0.742^{+0.185}_{-0.340}$ & $0.480^{+0.285}_{-0.290}$ \\
$\theta_1$ & $0.00$ & $0.288^{+0.139}_{-0.138}$ & $0.306^{+0.234}_{-0.163}$ & $0.355^{+0.189}_{-0.159}$ \\
$\theta_2$ & $0.00$ & $0.729^{+0.534}_{-0.400}$ & $0.362^{+0.345}_{-0.182}$ & $1.019^{+0.513}_{-0.476}$ \\
$d_L [Mpc]$ & $3000.00$ & $2921.335^{+289.478}_{-292.267}$ & $2762.648^{+293.087}_{-338.570}$ & $2800.689^{+283.210}_{-272.786}$ \\
$\theta_{jn}$ & $0.40$ & $0.387^{+0.156}_{-0.141}$ & $0.466^{+0.253}_{-0.224}$ & $0.363^{+0.140}_{-0.136}$ \\
$\phi$ & $1.30$ & $3.303^{+0.361}_{-0.393}$ & $1.292^{+3.987}_{-0.763}$ & $3.231^{+0.318}_{-0.376}$ \\
\bottomrule

\toprule
Parameter & Truth & IMRPhenomXPHM & IMRPhenomTPHM & IMRPhenomXO4a \\
\midrule
$\mathcal{M}_c [M_{\odot}]$ & $100.00$ & $98.640^{+4.418}_{-7.512}$ & $96.693^{+3.477}_{-6.154}$ & $90.705^{+3.341}_{-3.556}$ \\
q & $0.75$ & $0.601^{+0.175}_{-0.153}$ & $0.665^{+0.207}_{-0.175}$ & $0.478^{+0.059}_{-0.062}$ \\
$a_1$ & $0.75$ & $0.884^{+0.070}_{-0.088}$ & $0.815^{+0.107}_{-0.113}$ & $0.848^{+0.074}_{-0.065}$ \\
$a_2$ & $0.75$ & $0.609^{+0.234}_{-0.282}$ & $0.651^{+0.250}_{-0.354}$ & $0.550^{+0.247}_{-0.278}$ \\
$\theta_1$ & $0.00$ & $0.248^{+0.138}_{-0.124}$ & $0.303^{+0.200}_{-0.158}$ & $0.371^{+0.133}_{-0.153}$ \\
$\theta_2$ & $0.00$ & $0.735^{+0.614}_{-0.391}$ & $0.417^{+0.484}_{-0.227}$ & $1.256^{+0.469}_{-0.538}$ \\
$d_L [Mpc]$ & $3000.00$ & $2924.417^{+296.176}_{-291.803}$ & $2783.886^{+248.299}_{-283.127}$ & $2728.599^{+212.920}_{-219.523}$ \\
$\theta_{jn}$ & $0.40$ & $0.253^{+0.155}_{-0.115}$ & $0.266^{+0.180}_{-0.124}$ & $0.287^{+0.110}_{-0.108}$ \\
$\phi$ & $1.30$ & $5.368^{+0.619}_{-1.825}$ & $2.102^{+3.096}_{-1.225}$ & $4.948^{+0.522}_{-0.705}$ \\
\bottomrule

\toprule
Parameter & Truth & IMRPhenomXPHM & IMRPhenomTPHM & IMRPhenomXO4a \\
\midrule
$\mathcal{M}_c [M_{\odot}]$ & $100.00$ & $96.023^{+2.983}_{-5.080}$ & $98.996^{+2.416}_{-3.976}$ & $97.348^{+1.791}_{-2.049}$ \\
q & $0.75$ & $0.586^{+0.086}_{-0.118}$ & $0.660^{+0.118}_{-0.119}$ & $0.676^{+0.044}_{-0.047}$ \\
$a_1$ & $0.90$ & $0.916^{+0.038}_{-0.044}$ & $0.929^{+0.039}_{-0.047}$ & $0.924^{+0.043}_{-0.054}$ \\
$a_2$ & $0.90$ & $0.896^{+0.066}_{-0.126}$ & $0.919^{+0.052}_{-0.097}$ & $0.924^{+0.048}_{-0.093}$ \\
$\theta_1$ & $0.00$ & $0.252^{+0.113}_{-0.115}$ & $0.188^{+0.122}_{-0.098}$ & $0.278^{+0.130}_{-0.132}$ \\
$\theta_2$ & $0.00$ & $0.353^{+0.291}_{-0.188}$ & $0.181^{+0.140}_{-0.097}$ & $0.306^{+0.191}_{-0.152}$ \\
$d_L [Mpc]$ & $3000.00$ & $2681.198^{+252.595}_{-266.889}$ & $2719.508^{+217.552}_{-226.138}$ & $2639.487^{+313.254}_{-334.077}$ \\
$\theta_{jn}$ & $0.40$ & $0.422^{+0.190}_{-0.169}$ & $0.453^{+0.188}_{-0.171}$ & $0.617^{+0.158}_{-0.176}$ \\
$\phi$ & $1.30$ & $0.788^{+3.879}_{-0.399}$ & $1.743^{+3.470}_{-1.026}$ & $0.804^{+0.227}_{-0.230}$ \\
\bottomrule

\toprule
Parameter & Truth & IMRPhenomXPHM & IMRPhenomTPHM & IMRPhenomXO4a \\
\midrule
$\mathcal{M}_c [M_{\odot}]$ & $100.00$ & $99.092^{+3.522}_{-5.345}$ & $97.358^{+3.649}_{-4.772}$ & $97.003^{+3.053}_{-3.959}$ \\
q & $0.75$ & $0.537^{+0.118}_{-0.093}$ & $0.554^{+0.135}_{-0.106}$ & $0.508^{+0.080}_{-0.069}$ \\
$a_1$ & $0.75$ & $0.900^{+0.059}_{-0.070}$ & $0.828^{+0.082}_{-0.102}$ & $0.904^{+0.048}_{-0.062}$ \\
$a_2$ & $0.25$ & $0.268^{+0.351}_{-0.199}$ & $0.341^{+0.365}_{-0.247}$ & $0.544^{+0.279}_{-0.330}$ \\
$\theta_1$ & $0.20$ & $0.188^{+0.103}_{-0.087}$ & $0.450^{+0.208}_{-0.181}$ & $0.331^{+0.156}_{-0.151}$ \\
$\theta_2$ & $0.80$ & $2.148^{+0.563}_{-0.995}$ & $1.169^{+0.862}_{-0.615}$ & $2.074^{+0.447}_{-0.439}$ \\
$d_L [Mpc]$ & $3000.00$ & $2952.362^{+248.678}_{-260.318}$ & $2674.372^{+220.181}_{-217.242}$ & $2802.665^{+276.841}_{-235.559}$ \\
$\theta_{jn}$ & $0.40$ & $0.232^{+0.127}_{-0.101}$ & $0.337^{+0.131}_{-0.116}$ & $0.252^{+0.098}_{-0.090}$ \\
$\phi$ & $1.30$ & $2.456^{+0.559}_{-0.652}$ & $2.671^{+2.839}_{-2.062}$ & $2.375^{+0.442}_{-0.556}$ \\
\bottomrule

\end{tabular}

\caption{In this table, we show the waveform systematics effects associated to the merger of heavy mass BBHs, by showing the deviations in the estimations of the source parameters in the high-spin regime. We inject the signal with \texttt{NRSur7dq4} waveform-model and recover the source properties using \texttt{IMRPhenomXPHM-SpinTaylor}, \texttt{IMRPhenomTPHM} and \texttt{IMRPhenomXO4a} models. The first three parts of the table capture the systematics in the high spin regime $a_1, a_2 \geq 0.5$, with aligned spins. However, the fourth and final part also shows a misaligned binary with high-spins. The injected values of the spins and spin-angles are mentioned in the first column of the table. The distribution shapes and the correlation between different source parameters can be found in figure \ref{fig:spin_0.9} ($a_1 =a_2 = 0.9$ with aligned spins) and figure \ref{fig:spin_0.75_mis} ($a_1 = 0.75, a_2 = 0.25$ with misaligned spins).}
\label{tab:spin_posteriors_high_spin}
\end{table}

In the main text, we tried to decipher the the false lensing alarm for GW231123 due to the short duration of the signal in the observable frequency-band. The duration of the signal relies on the chirp mass of the binary system, controlling the frequency evolution of the emitted GW. Here, we check the false-lensing alarm of GW231123-like events due to the spin angular momentum of the components of the BBH system. We choose a binary system of chirp mass, $\mathcal{M}_c = 100 M_{\odot}$, a mass ratio of $q=0.75$ and place the source at a distance of $d_L = 3 \text{Gpc}$. This is to ensure that the network-SNR of the detected signal (with H1 and L1) is similar to the observed matched-filter SNR of GW231123. The GW waveforms are generated using \texttt{NRSur7dq4} waveform, which produced the least residual cross-correlation and the source parameters are recovered using the waveform models \texttt{IMRPhenomXPHM-SpinTaylor}, \texttt{IMRPhenomTPHM} and \texttt{IMRPhenomXO4a}. We divide the results into two parts, where the spin magnitudes are less than 0.5, i.e. $a_1, a_2 \leq 0.5$ in table \ref{tab:spin_posteriors_low_spin} and where the spin magnitudes are greater than 0.5, i.e. $a_1, a_2 \geq 0.5$ in table \ref{tab:spin_posteriors_high_spin}.

\begin{figure}
    \centering
    \includegraphics[width=\linewidth]{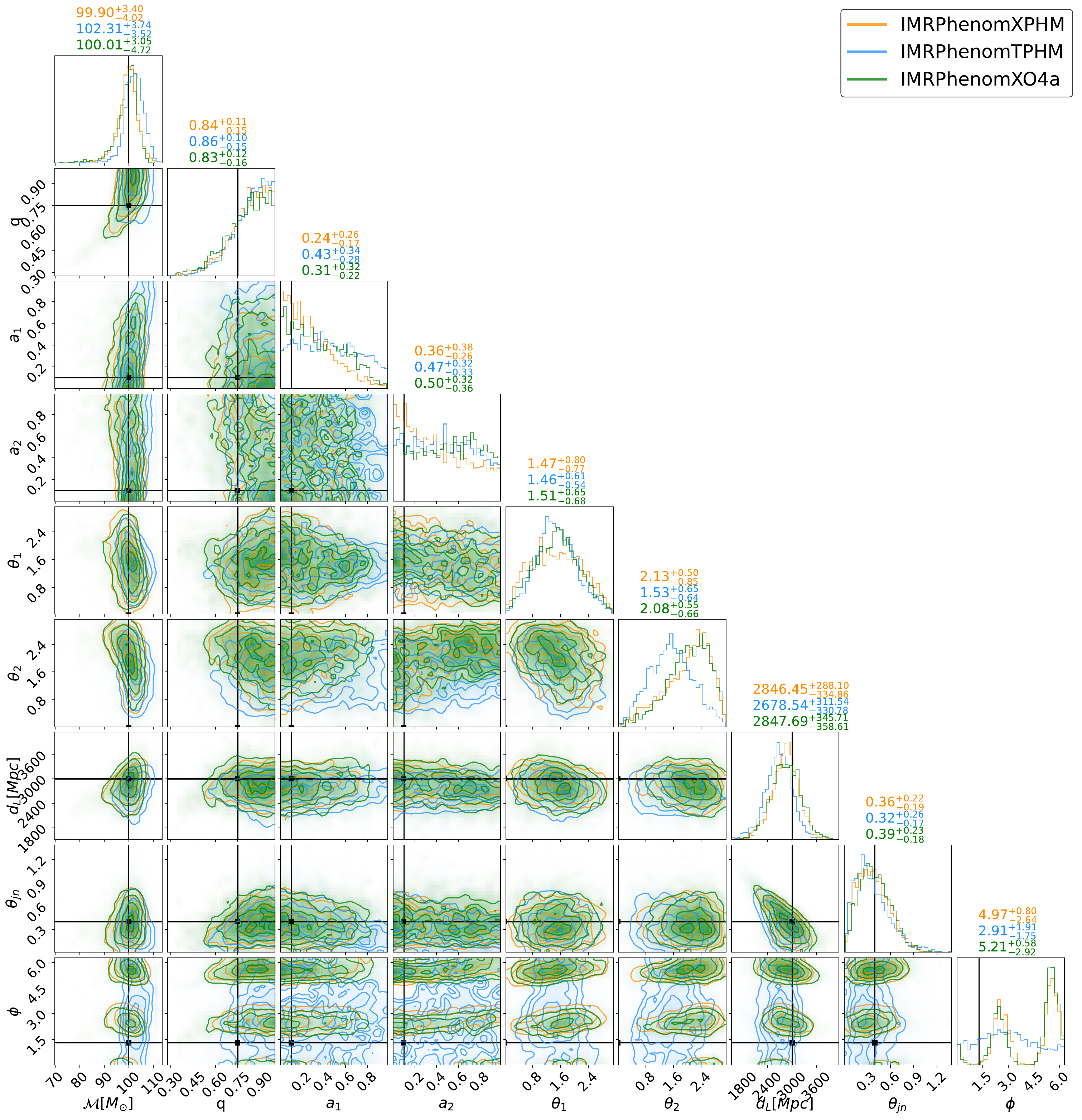}
    \caption{Source parameter joint distributions for $a_1 =a_2 = 0.1$ with aligned spins along the orbital angular momentum. A summary form of its important statistical quantities can be found in the table \ref{tab:spin_posteriors_low_spin}}
    \label{fig:spin_0.1}
\end{figure}

\begin{figure}
    \centering
    \includegraphics[width=\linewidth]{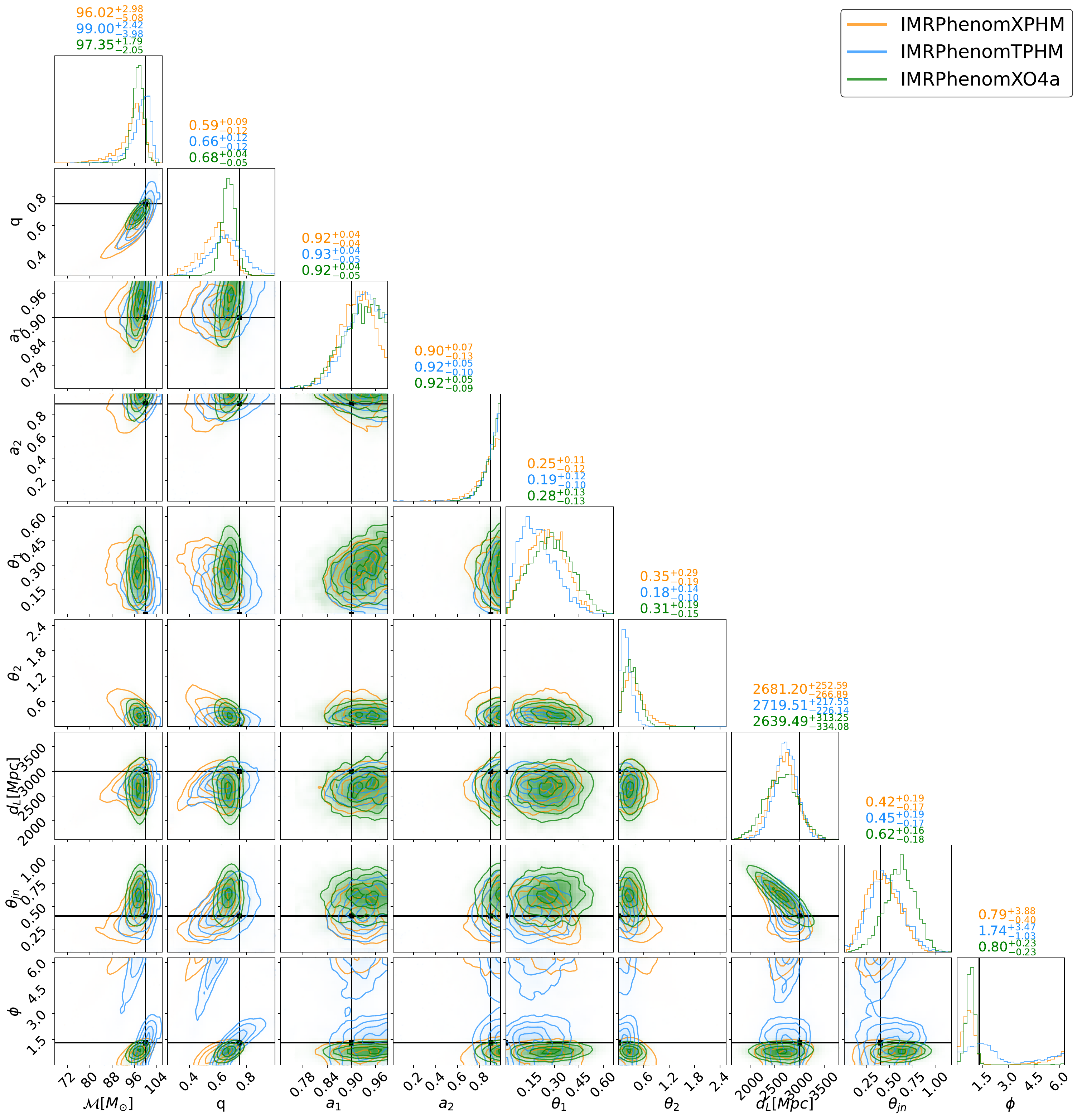}
    \caption{Source parameter joint distributions for $a_1 =a_2 = 0.9$ with aligned spins along the orbital angular momentum. A summary form of its important statistical quantities can be found in the table \ref{tab:spin_posteriors_high_spin}}
    \label{fig:spin_0.9}
\end{figure}

\begin{figure}
    \centering
    \includegraphics[width=\linewidth]{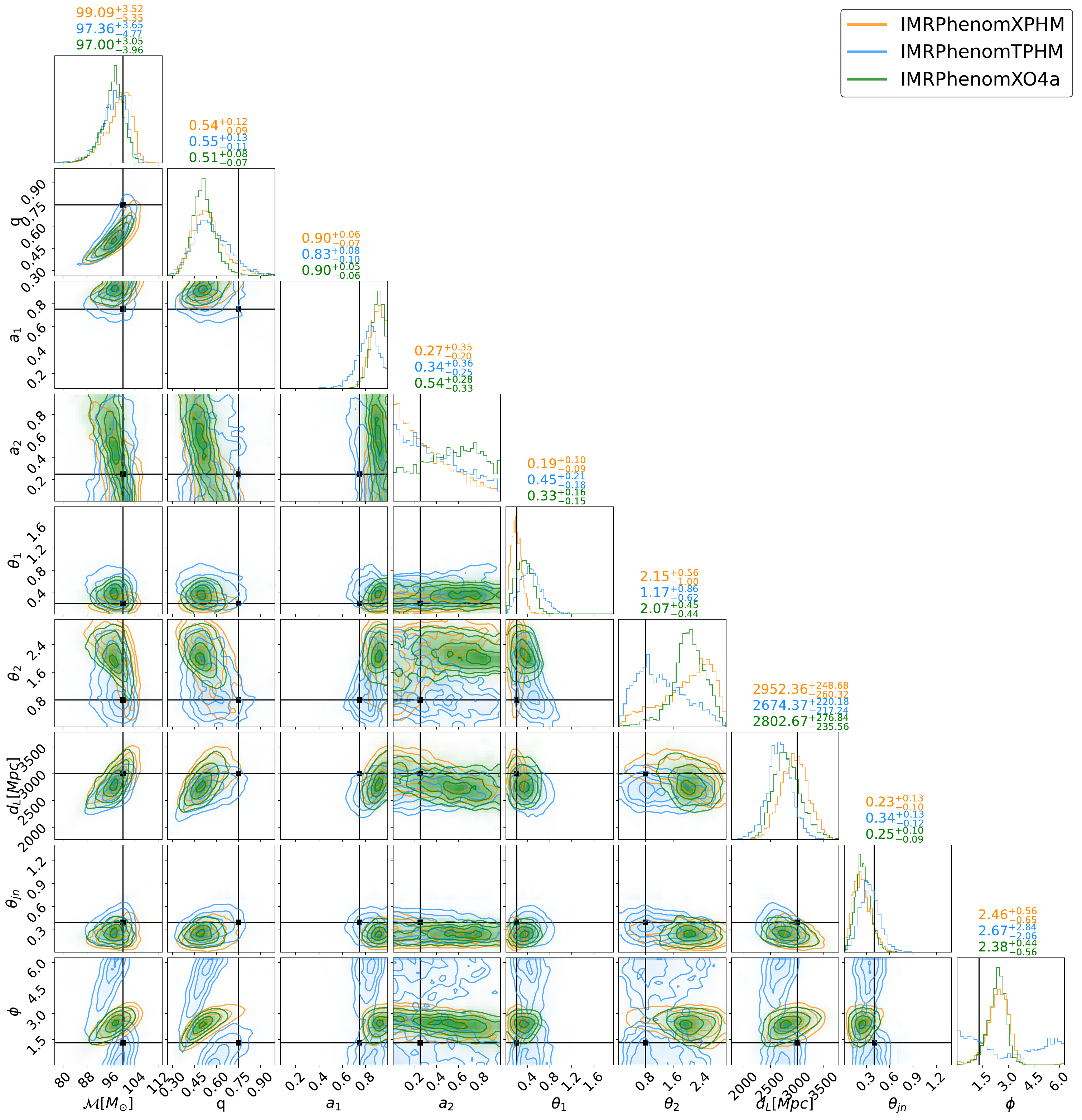}
    \caption{Source parameter joint distributions for $a_1 = 0.75, a_2 = 0.25$ with misaligned spins off the orbital angular momentum with spin angles $\theta_1 = 0.2$ and $\theta_2=0.8$. A summary form of its important statistical quantities can be found in the table \ref{tab:spin_posteriors_high_spin}}
    \label{fig:spin_0.75_mis}
\end{figure}

We observe that the chirp mass estimates are not significantly affected much for any of the low-spin or high-spin cases, the median of the distributions lie within the 3-$\sigma$ of the injected value of $\mathcal{M}_c = 100M_{\odot}$. However, we do note that, although the mass-ratio medians are consistently lying within the 3-$\sigma$ of the injected value of $q=0.75$ for the the different low-spin cases, this behaviour does not hold in the high-spin regime. For the case, $a_1 = a_2 =0.75 $, in the IMRPhenomXO4a recovery, the injected mass-ratio is at $\approx 5\sigma$ away from the median. This exact behaviour is also noticed for the misaligned spin-systems with spins $a_1 = 0.75, a_2 =0.25 $ and spin-angles $\theta_1 =0.2$ and $\theta_2 = 0.8$ system for the IMRPhenomXO4a recovery. For the the spin magnitudes, the posteriors agree with the injections up to the $3\sigma$ mark although the spin posteriors are broad and thus put weak constraints on the magnitudes for both low-spin and high-spin cases. Spin angles are not measured well. However, distance and inclination angles are estimated correctly with the different waveforms. Phase angles are again very poorly constrained. Marginalized 1D and 2D distributions of the source parameters for three boundary cases $a_1= a_2 =0.1$ (aligned), $a_1 = a_2 =0.9$ (aligned) and $a_1 = 0.75, a_2 =0.25$ (misaligned) are presented in the figure \ref{fig:spin_0.1}, figure \ref{fig:spin_0.9} and figure \ref{fig:spin_0.75_mis} respectively.

To summarize, these results point out that, the chirp-mass estimates, which govern the frequency evolution, are not affected significantly by BH spins. However, the mass-ratio, which decides higher order correction to the GW phase estimates can be impacted significantly. There are also previous studies which point out the mass-ratio and spin degeneracy \citep{Adamcewicz:2022hce, Adamcewicz:2023mov, Banerjee:2024wbq, Kang:2025nio}.
Thus, it is evident waveform systematics can play a significant role in collecting the residual of a GW event and in attributing the residual to a lensing modulation effect.

\end{document}